\documentclass[hidelinks,12pt]{article}
\usepackage[letterpaper,margin=1.25in]{geometry}

\usepackage{setspace}
\usepackage[T1]{fontenc}
\usepackage[utf8]{inputenc}

\usepackage[sc]{mathpazo}

\usepackage{amsmath,amssymb,mathtools}
\usepackage{amsthm}
\usepackage{bm}
\usepackage{bbm}

\DeclareMathOperator*{\supp}{supp}

\newcommand{\up}[1]{\overline{#1}}
\newcommand{\dw}[1]{\underline{#1}}

\def\sym#1{\ifmmode^{#1}\else\(^{#1}\)\fi}

\newtheorem{theorem}{Theorem}
\newtheorem{proposition}{Proposition}

\newtheorem{corollary}{Corollary}

\newtheorem*{remark*}{Remark}

\usepackage{graphicx}
\graphicspath{{./graphs/}}

\usepackage[dvipsnames]{xcolor}

\usepackage{tikz}
\usetikzlibrary{patterns.meta}

\definecolor{first_best}{RGB}{75,130,180}
\definecolor{cream}{RGB}{245,235,210}
\definecolor{Redish}{HTML}{890F0F}
\definecolor{general_color}{HTML}{FF7F00}

\usepackage{booktabs}
\usepackage{multirow}
\usepackage{hhline}

\usepackage[skip=0.5em,font=footnotesize]{caption}
\usepackage{subcaption}

\usepackage{soul}
\usepackage{comment}
\usepackage[super]{nth}
\usepackage[nice]{nicefrac}
\usepackage{enumitem}
\usepackage{multicol}
\usepackage{wrapfig}

\usepackage{float}

\usepackage{xr-hyper}
\usepackage{hyperref}

\hypersetup{
    colorlinks=true,
    linkcolor=general_color,
    filecolor=general_color,
    urlcolor=general_color,
    citecolor=general_color
}

\newcommand*{\fullref}[1]{\hyperref[{#1}]{\ref{#1} \nameref{#1}}} 
\newcommand*{\figref}[1]{\hyperref[{#1}]{Figure \ref{#1}}}
\newcommand*{\secref}[1]{\hyperref[{#1}]{Section \ref{#1}}}
\newcommand*{\tabref}[1]{\hyperref[{#1}]{Table \ref{#1}}}
\newcommand*{\Equationref}[1]{\hyperref[{#1}]{Equation \ref{#1}}}
\newcommand*{\equationref}[1]{\hyperref[{#1}]{equation \ref{#1}}}
\newcommand*{\Propref}[1]{\hyperref[{#1}]{Proposition \ref{#1}}}
\newcommand*{\theoref}[1]{\hyperref[{#1}]{Theorem \ref{#1}}}
\newcommand*{\Defref}[1]{\hyperref[{#1}]{Definition \ref{#1}}}
\newcommand*{\lemmaref}[1]{\hyperref[{#1}]{Lemma \ref{#1}}}
\newcommand*{\corref}[1]{\hyperref[{#1}]{Corollary \ref{#1}}}
\newcommand*{\assref}[1]{\hyperref[{#1}]{Assumption \ref{#1}}}
\newcommand*{\remref}[1]{\hyperref[{#1}]{Remark \ref{#1}}}

\usepackage[authoryear,round]{natbib}

\usepackage{xargs}
\usepackage[colorinlistoftodos,prependcaption,textsize=tiny]{todonotes}

\newcommandx{\unsure}[2][1=]{%
  \todo[linecolor=red,backgroundcolor=red!25,bordercolor=red,#1]{#2}}

\newcommandx{\change}[2][1=]{%
  \todo[linecolor=blue,backgroundcolor=blue!25,bordercolor=blue,#1]{#2}}

\newcommandx{\info}[2][1=]{%
  \todo[linecolor=OliveGreen,backgroundcolor=OliveGreen!25,bordercolor=OliveGreen,#1]{#2}}

\newcommandx{\improvement}[2][1=]{%
  \todo[linecolor=Plum,backgroundcolor=Plum!25,bordercolor=Plum,#1]{#2}}

\newcommandx{\thiswillnotshow}[2][1=]{%
  \todo[disable,#1]{#2}}

\title{\textbf{Delegated Contracting}\footnote{We thank Elliot Lipnowski, Leslie Marx, Andrew McClellan, Harry Pei, Doron Ravid, and Ludvig Sinander for useful comments and suggestions. We also thank the seminar audiences at Arizona State, Duke, FSU, QMUL, NC State, Oxford, Stanford, UBC, UNC, UC Riverside, and Yale for helpful feedback.}
\vspace{1em} 
\\}
\author{ João Thereze\footnote{The Fuqua School of Business, Duke University--- \href{mailto:joao.thereze@duke.edu}{joao.thereze@duke.edu}.}  ~~~~~~~~~~~~ Udayan Vaidya\footnote{The Fuqua School of Business, Duke University--- \href{mailto:udayan.vaidya@duke.edu}{udayan.vaidya@duke.edu}.} }
\date{\today}

\begin{document}

\maketitle
\thispagestyle{empty}
\setlength{\abovedisplayskip}{1.5pt}
\setlength{\belowdisplayskip}{3pt}

\begin{abstract}
A principal contracts with an agent through an informed delegate. Although the principal cannot directly mediate the interaction, she can restrict the menus of contracts the delegate may offer. We characterize the outcomes implementable through delegated contracting: they are exactly those achievable by a centralized Bayesian mechanism that is dominant-strategy incentive compatible and ex-post individually rational for the agent. We use this result to identify the optimal contractual restrictions in several settings. First, an organization that procures through a budget-indulgent agency should grant full flexibility below an expected spending cap. Second, unlike centralized mechanisms, delegated contracting can never dissolve partnerships efficiently, highlighting a limit to delegated authority. Finally, a seller can entrust sales to an intermediary without revenue loss by combining a resale price agreement with a buyback policy. 
\end{abstract}

\newpage
\pagenumbering{arabic}


\section{Introduction}
Principals often rely on intermediaries to contract with downstream agents. 
Organizations procure from suppliers through expert agencies, governments provide coverage to consumers through private insurers, and manufacturers reach final buyers through retailers. 
Despite not interacting directly with the agents, principals exert influence by restricting which contracts intermediaries may offer.
Organizations impose procurement rules, governments regulate insurance menus, and manufacturers constrain retailers' pricing policies. 
We refer to these environments as instances of \textit{delegated contracting}.

The prevalence of delegated contracting raises a basic problem of institutional design: how can a principal shape outcomes when authority must be exercised indirectly? 
To answer this question, we characterize the outcomes achievable under delegation, and then show how this characterization can be used to derive optimal contractual restrictions. 
Conceptually, our approach identifies the principal’s loss from delegation relative to centralized design.
Practically, we provide a tractable framework for analyzing optimal delegated contracting across a wide range of environments.

We formalize delegated contracting within a general framework. In our model, a principal must contract with an agent through a delegate. At the contracting stage, the delegate offers the agent a menu of outcomes, from which the agent chooses one option or opts out. The space of outcomes is unrestricted and may include transfers among the parties. Both the delegate and the agent have private information, yet the principal cannot directly communicate with the agent. Instead, her only design instrument is to impose ex ante restrictions on the menus the delegate may offer. The delegate selects an admissible menu given his private information, and the agent’s choice determines the payoffs of the contracting parties. Our baseline model focuses on a private values environment, though the analysis extends to more general settings.

Our model admits several interpretations.
It can be viewed as an organizational hierarchy that limits the discretion of subordinates, as a government regulating a mechanism designer, or as a firm imposing vertical controls within a supply chain.
These interpretations highlight a common structure underlying many delegated contracting environments.

We begin by characterizing which social choice rules---mappings from the parties' private information to outcomes---can arise through delegated contracting. \theoref{theo: implementability} shows that a social choice rule is implementable through delegation if and only if it is implementable in a direct mechanism, subject to (i) Bayesian incentive compatibility and interim participation for the delegate and (ii) dominant-strategy incentive compatibility and ex-post participation for the downstream agent.
The result builds on a simple argument, in the spirit of the revelation principle.
The stronger incentive constraints on the agent's side reflect the sequentiality of the environment: the delegate's choice of menu may signal private information that the agent can respond to. 
Under private values, there is no loss in allowing the delegate's menu to fully reveal his private information. 
As a result, incentive compatibility for the agent must hold uniformly across the delegate's types.

\theoref{theo: implementability} provides a useful organizing principle for studying delegated contracting. 
In particular, it recasts the principal's design problem---choosing a set of allowable menus---as a standard mechanism design problem with additional constraints.
This formulation opens the door to applying existing tools and insights from the broader mechanism design literature to delegated contracting environments, which we exploit in the applications below.
Moreover, \theoref{theo: implementability} identifies the cost of delegation relative to centralized contracting as the gap between dominant-strategy and Bayesian incentive constraints for the agent.

Our implementability result extends beyond the private values case, and we discuss two extensions in \secref{sec:discussionofassumptions}. First, we consider interdependent values under the assumption that the delegate is less informed than the agent about common payoff-relevant variables. In this case, our implementation result holds verbatim. Second, we extend our implementability result to an environment where one delegate interacts simultaneously with multiple agents, as in an auction.

\paragraph{Applications.}
We leverage \theoref{theo: implementability}  to identify optimal contractual restrictions in several environments. Our applications provide practical insights about the trade-off between authority and flexibility in delegated contracting. Moreover, these environments clarify when the loss from delegation is negligible or substantial.

First, we study a procurement setting in which an organization (the principal) relies on a division (the delegate) to procure goods from a supplier (the agent).
The division is privately informed about the benefit of procurement to the organization, but is budget-indulgent. The division screens the supplier, whose cost is privately known, by offering a menu of price-quantity pairs. We show that a version of interval delegation is optimal for the organization: the division is allowed to offer its preferred menu to the supplier, as long as its benefit level lies below a cap. 
Under additional assumptions on the supplier's cost, it is sufficient for the organization to restrict only the procurement \emph{outcomes} themselves, rather than the menus offered. Concretely, the organization specifies a maximum price for each level of quantity, and the division is free to offer any menu lying below this price-quantity frontier. These results extend the typical logic of interval delegation \citep{alonso_optimal_2008, amador_regulating_2022} to our environment, where the delegate's action space is rich. Our proof technique allows these results to readily generalize to more complex procurement problems, including those with multiple competing suppliers.

In our second application, we ask whether delegated contracting obstructs the implementation of efficient outcomes. 
We focus on a canonical allocation problem, where the principal assigns an asset to one of two players under budget balance. The only departure from the standard setting is that the principal must contract with one of the players (the agent) through the other (the delegate).
When the players have no initial ownership of the asset being allocated, we show there is no loss from delegation. 
However, when the delegate and agent jointly own the whole asset, delegation precludes efficiency regardless of the ownership split.
This result highlights a stark gap between centralized and delegated contracting. While a centralized mechanism can always implement efficiency when ownership shares are sufficiently similar \citep{cramton_dissolving_1987}, delegated implementation is never efficient.

We then study how the allocation of contracting rights can facilitate efficiency, that is, whom to delegate to. We characterize the subsidy-minimizing delegate when the players differ either in terms of their private information or their initial ownership of the asset. In both cases, the efficient outcome is implemented through a menu of bid-ask pairs, linking our results to the design of buy-sell clauses, mechanisms used to dissolve partnership deadlocks. In particular, we show that separating the buy and sell prices, together with identifying the correct proposer, can enhance efficiency. Taken together, our analyses underscore the role of mediation in efficiently reallocating ownership and highlight the importance of jointly studying property rights and contracting rights.


Our final application examines a vertical supply chain in which an intermediary (the delegate) facilitates trade between a seller (the principal) and a consumer (the agent).
The intermediary can either resell the good or use it for private benefit. 
We show that the seller can achieve the same profit through delegation as in an optimal auction involving the intermediary and the consumer. 
The corresponding contracting space effectively treats ownership and resale rights as two distinct objects. 
One simple implementation involves resale price maintenance and a buyback policy. 
Resale price maintenance induces the intermediary to transfer the good efficiently by tying the resale price to the seller's wholesale price. 
The buyback policy encourages intermediation because low-value intermediaries would not buy the good otherwise. 
The vertical controls imposed through delegated contracting increase the seller's and consumer's surplus relative to a \emph{laissez-faire} benchmark, at the expense of the intermediary.


\paragraph{Related Literature.}
A substantial literature examines the distortions that arise when a principal anticipates subcontracting among agents.\footnote{One focus of this line of work is studying ``collusion'' due to unobserved subcontracting \citep{tirole_hierarchies_1986,tirole_collusion_1992,mcafee_organizational_1995,laffont_collusion_1997,baliga_decentralization_1998,severinov_optimal_2003,mookherjee_organization_2004,mookherjee_decentralization_2006,mookherjee_incentives_2012}. In our setting, downstream contracting is observable and contractible, so the principal is not concerned with collusion \emph{per se}.} 
Several of these papers consider environments in which a delegated implementation replicates the optimal centralized mechanism \citep{melumad_communication_1991,melumad_theory_1992,melumad_hierarchical_1995,faure-grimaud_agency_2001,mitchell_pricing_2025, BhaskarMcClellan2025}.  Closest to us, \citet{bhaskar_regulation_2023} are the first to study regulation as contract-menu delegation. In an insurance setting, they show how requiring menus to include latent contracts allows the regulator to replicate a centralized benchmark.
Our paper characterizes exactly when centralized outcomes are implementable through menu restrictions and, when they are not, offers a program to solve for optimal delegation. 
When the principal cannot restrict menus directly, weaker contracting instruments include observing individual outputs \citep{mookherjee_incentives_2012}, constraining outcomes \citep{malladi_delegated_2022}, or designing players' outside options \citep{dworczak_mechanism-design_2024}. In these cases, delegation limits implementation.
By contrast, \citet{basford2026commitment} show that, in a bilateral trade setting, ex ante commitment by one player to a third-party contract expands the set of implementable outcomes.

We also relate to the literature on informed mechanism design \citep{myerson1983mechanism, maskin1990principal, mylovanov2014mechanism}, as the delegate in our model can be seen as a privately informed designer. Unlike the typical mechanism selection game, our principal limits the set of feasible mechanisms, and need not share the delegate's preferences. Moreover, the delegate is limited to offering menus of contracts to the agent. This is a restriction to \emph{transparent} mechanisms \citep{zheng_optimal_2002}, rather than arbitrary general mechanisms, where the inscrutability principle applies. This restriction captures the constraints imposed by delegation relative to mediation, and is natural in many settings. We discuss transparency in \secref{sec:discussionofassumptions}.

Finally, we build on the delegation literature initiated by \citet{holmstrom_theory_1980}. In standard one-dimensional settings, interval delegation is often optimal \citep{alonso_optimal_2008,amador_theory_2013,amador_regulating_2022,guo_dynamic_2016,kolotilin_persuasion_2025}. In richer environments, however, optimal delegation sets may be more complex \citep{ambrus_delegation_2017,koessler_optimal_2012,frankel_aligned_2014,kleiner2025optimal}. Our procurement application extends the interval-delegation logic to a setting in which the delegate offers a menu of contracts to a privately informed agent. The proof does so by replacing the agent's incentive constraints with an upper bound on the delegate's payoff, a relaxation related to \citet{kleiner2025optimal}. Lastly, unlike regulatory models that restrict attention to particular constrained delegation mechanisms \citep{kundu_delegation_2020,martimort_use_2020}, our framework characterizes the optimal contracting restrictions directly.


\section{A model of delegated contracting} \label{sec: model}

A principal (she) seeks to achieve an outcome through the interaction between two privately informed players, indexed by $i \in \{1,2\}$. We refer to player 1 as the \textit{delegate} and player 2 as the \textit{agent}. Each player $i$ holds private information or type $\theta_i$, with $(\theta_1,\theta_2)$ drawn jointly from  $\Theta := \Theta_1 \times\Theta_2$, according to a probability measure $\mu_o \in \Delta\Theta$. The space of outcomes is denoted by $X$, and player $i$'s payoff from outcome $x \in X$ given private information $\theta_i$ is $u_i(x, \theta_i) \in \mathbb{R}$. If no agreement is reached, a default outcome $o \in X$ occurs.\footnote{The sets $\Theta_i$ and $X$ are closed subsets of finite-dimensional Euclidean spaces, each endowed with its respective Borel sigma-algebra. Furthermore, we endow the power set of $X$, $2^X$, with a sigma-algebra $\Sigma$ that includes all singletons. That is, if $C \in 2^X$, then $\{C\} \in \Sigma$. Finally, the payoff functions $u_i$ are measurable.} 

The principal cannot mediate the interaction between the delegate and the agent. Instead, she entrusts contractual authority to the delegate, who interacts privately with the agent. Yet, the principal retains control over the contract space by specifying the allowable menus of contracts. 
Formally, a \textit{contract} is simply an outcome $x \in X$ agreed upon by the delegate and the agent, e.g., a price-quantity pair. In turn, a menu is a non-empty set of contracts, $C \subseteq X $. The timing of the game is as follows. (1) the principal specifies the delegate's contracting rights, $\mathcal{C} \subseteq 2^X$. Contracting rights are a collection of subsets of $X$, representing the admissible menus of contracts that can be offered to the agent. (2) The delegate observes his type $\theta_1$ and either selects an admissible menu $C \in \mathcal{C}$ to offer or opts out, yielding outcome $o$. (3) The agent observes his type $\theta_2$ and chooses to sign an available contract $x \in C$ or to opt out, yielding outcome $o$.

We now define the strategies in the subgame between the delegate and agent, given the principal's choice of contracting rights $\mathcal{C}$. The delegate's strategy $\sigma_1: \Theta_1 \rightarrow \Delta 2^X$ maps the delegate's information to a lottery over admissible menus. That is, $\sigma_1(\theta_1)$ almost surely selects a menu in $\mathcal{C} \cup \{\{o\}\}$.\footnote{If $\mathcal{C}$ is measurable, this requires $\sigma_1(\theta_1)(\mathcal{C} \cup \{\{o\}\})=1$, as usual. If $\mathcal{C}$ is not measurable, we instead require that the inner measure of $\mathcal{C}$ is one. Formally: $$\sup_{\substack{A \in \Sigma \\ A\subseteq \mathcal{C}\cup \{\{o\}\}}} \sigma_1(\theta_1)(A)=1.$$ By the assumption that $\Sigma$ includes all singletons, a pure strategy by the delegate is always allowed. } Similarly, the agent's strategy given the menu $C$, $\sigma_2^C: \Theta_2 \rightarrow \Delta X$, maps the agent's information to a lottery over included contracts. That is, $\sigma_2^C(\theta_2)$ almost surely selects an outcome in $C\cup\{o\}$. The agent's beliefs over the delegate’s type after observing the menu $C$ are given by $\mu^C: \Theta_2 \rightarrow \Delta \Theta_1$. 
\subsection{Delegated Implementation}
Our first goal is to establish what outcomes can be obtained through delegated contracting. Formally, a social choice function is a measurable mapping between the private information of both players and the set of outcomes, $Y: \Theta \rightarrow X$.\footnote{The assumption that the social choice function is deterministic is without loss of generality in our model, up to measurability restrictions. See \secref{sec:discussionofassumptions}.} A social choice function is \emph{implementable through delegation} if there exist contracting rights $\mathcal{C}$ and an associated perfect Bayesian equilibrium (henceforth, equilibrium) $\left(\sigma_1, \{\sigma_2^C,\mu^C\}_{C \in \mathcal{C}}\right)$ inducing outcomes given by $Y$. That is, given the equilibrium strategies, for any types $(\theta_1,\theta_2) \in \supp \mu_o$, outcome $Y(\theta_1,\theta_2)$ arises almost surely:

$$\int_C \sigma^C_2 (\theta_2) \left( Y(\theta_1,\theta_2) \right) \sigma_1 (\theta_1) (dC) = 1.  $$

We compare the outcomes implementable through delegation with those achievable via \textit{centralized mechanisms}, in which the principal directly communicates with both parties. A centralized mechanism is a pair $(M, g)$ with $M = M_1 \times M_2$ and $g: M \rightarrow X$, where $M$ is a message space and $g$ is an outcome function that maps messages into a chosen contract. The direct mechanism associated with $Y$ is one with $M = \Theta$, and $g=Y$. 

We start by restating standard definitions. We say a social choice function is Bayesian incentive compatible (BIC) for player $i$ if its corresponding direct mechanism induces truthful reporting given $i$'s beliefs about the other player's type.  It is interim individually rational (IIR) if $i$ prefers to participate in this mechanism rather than to obtain the disagreement outcome. Formally, a social choice function $Y$ is BIC and IIR for player $i$ if for all $\theta_i, \theta' \in \Theta_i$:
$$ \mathbb{E}_{\theta_{-i}}\left[u_i\left(Y(\theta_i, \theta_{-i}), \theta_i \right) | \theta_i\right] \geq \max\left\{\mathbb{E}_{\theta_{-i}}\left[u_i\left(Y(\theta', \theta_{-i}), \theta_i \right) | \theta_i\right], u(o,\theta_i) \right\}.  $$

A social choice function is said to be dominant-strategy incentive compatible (DSIC) and ex-post individually rational (EPIR) for party $i$ if in the corresponding direct mechanism $i$ prefers to report truthfully and participate knowing the other player's type. Formally, for all $\theta_i, \theta' \in \Theta_i$, and $\theta_{-i} \in \Theta_{-i}$:
$$ u_i\left(Y(\theta_i, \theta_{-i}), \theta_i \right) \geq \max\left\{u_i\left(Y(\theta', \theta_{-i}), \theta_i \right), u(o,\theta_i) \right\}.  $$

Our main result characterizes delegated implementation in terms of these standard incentive constraints. In the centralized case, the revelation principle asserts that a social choice function is Bayes implementable if and only if it is BIC and IIR for both the delegate and the agent \citep{myerson_optimal_1981}. We show that delegated contracting narrows the set of outcomes that the principal can achieve.

\begin{theorem} \label{theo: implementability}
A social choice function $Y$ is implementable through delegation if and only if it satisfies:
\begin{enumerate}[label=(\roman*)]
    \item Bayesian incentive compatibility (BIC) and interim individual rationality (IIR) for the delegate, and
    \item Dominant-strategy incentive compatibility (DSIC) and ex-post individual rationality (EPIR) for the agent.
\end{enumerate}
\end{theorem}

\theoref{theo: implementability} shows that the principal can implement any outcome achievable under centralization, subject to one caveat: the agent must be willing to participate and report truthfully even after learning the delegate's type. The proof builds on a simple argument in the spirit of the revelation principle. If the social choice function is responsive to the delegate's type, his choice of menu signals his private information. This allows the agent to respond differently to different delegate types. Thus, Bayesian incentives are insufficient. Because of private values and menu contracting, there is no loss in the delegate's menu fully revealing his type, requiring agent incentives to hold type-by-type of the delegate.

This result is useful for two reasons. First, it provides a practical tool to study delegated contracting problems. Instead of solving the interaction between the delegate and the agent and proceeding by backward induction to search over the unstructured space of contracting rights, \theoref{theo: implementability} translates delegated contracting into a set of familiar implementation constraints. Second, the result precisely delineates the loss from delegation compared to standard, centralized mechanism design. In doing so, \theoref{theo: implementability} complements a strand of the literature that identifies environments where there is no loss from delegated contracting \citep{bhaskar_regulation_2023, BhaskarMcClellan2025, mitchell_pricing_2025}, but does not quantify the extent of such losses when they arise.


\subsection{Discussion of Assumptions} \label{sec:discussionofassumptions}
\paragraph{Menu Contracting vs. General Mechanisms} A key assumption in our model is that the delegate contracts by offering a menu of outcomes. Unlike general mechanisms, which can depend on reports from both players, menu contracting restricts the delegate to offer \emph{transparent} mechanisms \citep{zheng_optimal_2002}. This restriction imposes that the informed delegate cannot affect the outcome beyond the choice of menu.   

Transparency is appealing for several reasons. First, it captures the constraints imposed by delegation. Given the principal’s inability to monitor the interaction, transparency ensures that the agent himself can audit the delegate. Any potential deviation by the delegate would be evident to the agent because he directly chooses an outcome. 
Second, transparent mechanisms are strategically simple \citep{borgers2019strategically}, requiring players to form only first-order beliefs about each other. Strategic simplicity is a desirable property when the principal is not confident about the players' higher-order beliefs. Simplicity is especially relevant when the agent may be less cognitively sophisticated than the delegate, for example when the principal regulates the relationship between a large firm and an individual consumer.   Finally, menu-like contracting processes are ubiquitous in practice.

If general mechanisms were allowed, our problem would reduce to a standard one. The principal could implement any outcome attainable in a centralized Bayesian mechanism by delegating a single inscrutable mechanism \citep{myerson1983mechanism}.

\paragraph{Private Values} The private values assumption implies that the agent's preferences are independent of the delegate's information, ruling out any non-trivial signaling by the delegate's choice of menu. Nonetheless, \theoref{theo: implementability} extends verbatim to certain interdependent value environments where the agent is better-informed about his payoff-relevant states than the delegate. We formalize this result in Appendix \ref{sec: extensions}, \theoref{theo: implementability with interdependence}. These environments include ``one-directional'' interdependent values models, where the delegate's preferences depend on the agent's type, but not vice versa. This is typical in problems of screening under adverse selection, including \citet{bhaskar_regulation_2023}. For instance, while a  lender might be wary of a borrower's risk type, the borrower need not contemplate the lender's costs given the loan's interest rate.

\paragraph{Single downstream agent.} Our model assumes the delegate only interacts with a single downstream agent, though there are many settings where it is natural for there to be multiple downstream agents. For example, consider an artist selling through an auction house using a prespecified auction format. We extend \theoref{theo: implementability} to encompass these settings in Appendix~\ref{sec: extensions}, \theoref{theo: implementability with multiple agents}. Our multi-agent setup requires two small modifications. We assume that in the contracting stage, the delegate can use any (transparent) mechanism whose outcome depends only on the agents' reports. In addition, we require the delegate's type to be independent of the agents' types. If the types were correlated, the delegate's choice of mechanism would allow agents to update about each others' types, tightening their incentive constraints. Under these assumptions, we obtain the natural generalization of \theoref{theo: implementability}. In this characterization, the downstream agents must have dominant strategy incentives with respect to the delegate's type, but only Bayesian incentives with respect to other agents' types. 

\paragraph{Deterministic Social Choice Functions} To ease the setup, we assumed social choice functions are deterministic, but this is without loss of generality. In order to accommodate random social choice functions, one can redefine contracts as belonging to ${\Delta X}$. In this case, menus consist of non-empty sets of lotteries. Under the appropriate measurability assumptions, embedding randomization into the contracts this way lets \theoref{theo: implementability} apply to stochastic social choice functions.


\section{Applications}

\subsection{Delegated Procurement}

We apply our delegated contracting framework to study a stylized procurement environment. In this setting, a principal delegates the task of contracting with a supplier (the agent) to an internal division (the delegate). For example, a legislature may delegate procurement authority to an expert agency, or a corporation may delegate purchasing decisions to regional offices. Our delegated procurement environment features a familiar tradeoff: the delegate has superior information about the value of procurement, but may be overindulgent with the principal's resources. How should the principal discipline the procurement process without directly intervening in the downstream contracting?

The procurement task involves securing some quantity $q \in [0,1]$ of a good delivered by the agent. Importantly, we assume that there are no internal transfers between the principal and the delegate, so a contract is simply a pair $(q,t) \in X := [0,1] \times \mathbb{R}$, where $t$ is a transfer paid to the agent. As in our general model, the principal cannot contract directly with the agent, but can restrict the menus of contracts the delegate may offer. 

After learning the principal’s contractual restrictions, the delegate privately observes a benefit parameter $b \in [0,1]$, which captures the delegate's per-unit benefit of procurement. We assume $b$ is distributed according to a CDF $F$ that admits a positive and differentiable density $f>0$. The agent privately observes a cost type $s \in S$, which determines the cost of supplying $q$ units according to the function $c(s,q)$. The agent's type space $S$ is an arbitrary measurable space endowed with probability measure $\mu$. We assume $s$ is independent of $b$, and for most of this section we impose no structural assumptions on the distribution $\mu$ or on the agent's cost function. Instead, we assume directly that for each type $b \in [0,1]$, the delegate’s screening problem formulated below in \eqref{eq: del_proc_problem} admits a unique and bounded solution.\footnote{In particular, we do not require monotone hazard rates, single-crossing, or convexity. Typical specifications of the cost function imply the boundedness and uniqueness of the solution, including the one we present in the next subsection.}
When the delegate procures quantity $q$ at transfer $t$, payoffs of the principal, delegate, and agent are given, respectively, by
\[
u_P = v(b)q - t, 
\qquad 
u_D = bq - t, 
\qquad 
u_A = t - c(s,q).
\]

The payoffs reflect that, although the principal and the delegate both dislike spending money, they differ in their rate of substitution between spending and procurement.
Whereas the delegate is willing to spend up to $b$ for each additional unit, the principal is only willing to spend $v(b)\geq 0$. We assume $v$ is a differentiable function that satisfies two properties: (i) $b \geq v(b)$ for all $b$, with strict inequality for some $b$, and (ii) the wedge $b - v(b)$ is non-decreasing in $b$. Thus, delegates are prone to overspending and will accordingly not screen the downstream agent strongly enough. 

This kind of misalignment is common within organizations. The gap between $b$ and $v(b)$ may reflect failure by the delegate to internalize a shadow cost of spending the principal's resources, such as the principal's opportunity cost of allocating money to other divisions or the distortionary effects of taxes required to raise funds. The wedge may also represent regulatory capture, causing the delegate to prefer excess procurement due to kickbacks or other informal relationships with the supplier. In what follows, we study how the principal optimally mitigates the overspending generated by this misalignment.

\paragraph{The delegate’s unconstrained screening problem.}  
Before attacking the principal's problem, a useful benchmark is to study the delegate's unconstrained screening problem. To that end, fix a delegate type $b$ and suppose the principal chose not to restrict the downstream contracting at all. The delegate would  then design a (transparent) screening mechanism to maximize his expected payoff subject to incentive compatibility and participation of the supplier. Invoking the revelation principle, the delegate's problem can be written
\begin{align}
\max_{q(\cdot),\, t(\cdot)} \; &\mathbb{E}_{s \sim \mu}\big[\, b q(s) - t(s) \,\big] \label{eq: del_proc_problem} \\
s.t. \quad &t(s) - c(s,q(s)) \ge t(s') - c(s,q(s')) \quad \forall s,s', \tag{DSIC - Agent}\\
&t(s) - c(s,q(s)) \ge 0 \quad \forall s  \tag{EPIR - Agent}
\end{align}

We have assumed that $c$, $S$, and $\mu$ are such that this problem admits a bounded and unique (up to $\mu$-measure-zero) solution for each $b$. Let $(q_b^*(\cdot), t_b^*(\cdot))$ denote the resulting $b$-optimal direct mechanism. By the taxation principle, each such direct mechanism can be implemented using a menu of contracts $C_b \subseteq X$, where $C_b = \cup_{s \in S} \big(q_b^*(s), t_b^*(s)\big)$. We refer to $C_b$ as the \textbf{$b$-optimal menu}. 

\paragraph{The principal's problem.} With the delegate's problem in hand, we are now ready to turn to the principal's problem. Our formulation allows the principal to implement many kinds of restrictions including ex-post quantity caps, expected spending caps, or bounds on the per-unit spending. To identify the optimal contractual restrictions directly, we invoke Theorem \ref{theo: implementability} and solve the corresponding direct mechanism problem. The solution is best described by way of the delegate's optimal menus. 

\begin{proposition}
\label{prop:procurement}
   Suppose the delegate's information satisfies the monotone hazard rate condition, i.e., $\frac{1-F}{f}$ is decreasing. Then, there exists a cutoff $\hat{b} < 1$ such that the principal's payoff is maximized by delegating an interval of $b$-optimal menus $\left(C_b\right)_{b \in [0,\hat{b}]}$. 
\end{proposition}

\Propref{prop:procurement} highlights three key properties of optimal delegation that extend insights from canonical models to our setting. First, the delegate's allocation is capped. Indeed, under the optimal delegation set, delegate types below $\hat{b}$ choose their optimal menus as if they were unconstrained, while those above $\hat{b}$ pool at the maximally-allowed menu, $C_{\hat{b}}$. This menu cap extends the optimality of price caps in the standard delegation and monopoly regulation literatures to an environment with multidimensional actions \citep{amador_regulating_2022, guo_dynamic_2016}. Second, the menus of contracts delegates may choose from are undistorted. Every type of delegate ultimately implements some $b$-optimal menu, potentially associated with a lower type. To curb overspending, the principal simply limits the set of allowable delegate-optimal actions. This property also holds in standard delegation problems, but is typically a direct consequence of the structure of the one-dimensional action space and type space. Here, this conclusion relies on linearity of both the designer's and the delegate's preferences.\footnote{For a simple example of how the observed menus could be distorted, suppose the principal imposed an ex-post quantity cap of $\bar{q}$. If for each $b$, $\max_s q^*_b(s) > \bar{q}$, then no type of the delegate could use a $b$-optimal menu. For the comparison with standard delegation, consider a canonical one-player delegation model where actions are a subset of $[0,1]$ and delegates are identified by their ideal action. In such a model, so long as non-degenerate delegation occurs, every chosen action is optimal for some type. } Finally, delegation has an interval structure. The optimal delegation set is an interval of contracts, ordered by the delegate's types, extending a key observation in the delegation literature \citep{alonso_optimal_2008, kolotilin_persuasion_2025}. 

One partial intuition for \Propref{prop:procurement} is that delegate-optimal menus satisfy a cost-minimization property. Given any target expected quantity $\bar Q$, both the principal and the delegate prefer to procure $\bar Q$ in the least costly way. Consider the problem of minimizing expected transfers to the agent subject to incentive compatibility, participation, and the quantity constraint $\mathbb{E}_s[q(s)] = \bar Q$. The solution to the cost-minimization problem must also maximize $\mathbb{E}_s[\lambda q(s) - t(s)]$, where $\lambda$ is the Lagrange multiplier on the quantity constraint. This program is simply the unconstrained maximization problem of the type-$\lambda$ delegate, so cost minimization pushes the principal to not distort the menus themselves.

The remaining question is which $b$-optimal menus should be permitted. The answer relies on the following key observation: given the linear payoff structure,
all the relevant details of the downstream procurement problem can be summarized by each delegate's expected payoff $U^*(b) = \mathbb{E}_{s}[b q_b^*(s)  - t_b^*(s)]$ when unconstrained.  We proceed by considering the following relaxation of the principal's problem, where the agent's incentive constraints are replaced by an upper bound on the delegate's payoff---the payoff of the delegate if unconstrained. 
\begin{align}
    \max_{q(b,s),t(b,s)} &\mathbb{E}_{b,s} \left[ v(b) q(b,s) - t(b,s) \right] \\
    s.t. \quad &\text{Bayesian IC \& Interim IR - Delegate} \nonumber \\
    &\mathbb{E}_s[b q(b,s) - t(b,s)] \leq U^*(b) \quad \forall b\label{eq: del_proc_UB}
\end{align}

The relaxed problem yields a one-dimensional screening problem, with an additional majorization-like constraint on the delegate's payoff (\ref{eq: del_proc_UB}). This allows us to solve for the principal's optimal expected quantity $Q(b)$ and transfer $T(b)$ for each delegate type.  Under our assumptions, the optimal solution takes a cutoff form. The constraint binds up to some $\hat b$ and is slack thereafter. The value of the relaxed program can then be achieved by delegating an interval of $b$-optimal screening menus up to $\hat{b}$.

In a related paper, \citet{kleiner2025optimal} shows that incentive compatible mechanisms are characterized by an upper bound on the delegate's payoff, the counterpart of our \eqref{eq: del_proc_UB}. His model considers a multidimensional delegation environment with no downstream interaction. While his constraint arises from the delegate's incentives, ours arises from menu contracting. If our principal could design a centralized procurement mechanism, such a constraint would not be appropriate. Once we invoke \theoref{theo: implementability} and replace the agent's constraints with this upper bound, standard arguments reduce our problem to the kind studied in \citet{kleiner2025optimal}.

An advantage of framing the analysis in terms of the delegate's unconstrained payoffs is that the result is robust to alternative specifications of the downstream procurement problem. The supplier’s type space may be irregular, the cost function may be non-convex, or multiple suppliers may compete in an auction. In each case, the details of the screening problem are  fully summarized by $U^*(b)$, and our proof technique applies. Though the delegate's optimal mechanisms may be complicated, the structure of optimal delegation remains: the principal should delegate an interval of $b$-optimal mechanisms. 

\paragraph{Implementation.} 
In our general setup, the principal can implement her optimal allocation through several methods. The most direct and restrictive method would be to have the delegate directly choose any menu from $(C_b)_{b \in [0,\hat{b}]}$; however, more flexibility can be given. For one, the principal could allow the delegate to choose any menu of contracts $C$ whose expected total quantity or transfer is bounded by that of $C_{\hat{b}}$. Both implementations require the designer to constrain the set of \emph{menus} that can be used by the delegate. However, in many situations the principal may be able to restrict only what \emph{outcomes} can be chosen, and the delegate may construct his menus as he wishes \citep{malladi_delegated_2022}.

Under a natural specification of the procurement problem, we show that such \textit{outcome-based} restrictions can also be optimal for the principal. To obtain this result, we specialize our model as follows. The agent's type is given by $s \in [\underline{s},1]$ for some $\underline{s} \in (0,1)$, and the agent's cost from producing $q$ units satisfies $c(s,q) = s\ k(q)$. We assume the function $k$ is strictly increasing, strictly convex, and satisfies the Inada conditions: $\lim_{q \rightarrow 0} k'(q) = 0$ and $\lim_{q\rightarrow \infty} k'(q) = \infty$. Additionally, we assume that the agent's type $s$ is drawn according to a distribution with density $g>0$. The principal and delegate's payoffs are unchanged. Under this specification, let $\mathcal{C}^{\hat{b}} := (C_b)_{b \in [0,\hat{b}]}$ denote the solution to the principal's problem identified in \Propref{prop:procurement}.

Within this setup, we argue that a \emph{price-quantity} frontier generates the same outcomes as $\mathcal{C}^{\hat{b}}$ and is, therefore, optimal as well. This frontier fixes the maximum price the principal will pay for each quantity level, and allows the delegate to freely offer any collection of contracts below this bound. The frontier $\tau(q)$ is defined by the maximal transfer associated with quantity $q$ across all menus in $\mathcal{C}^{\hat{b}}$. Formally, let \[\tau(q) = \sup \{ t \mid (q',t) \in C \text{ for some } q'\leq q \ \text{ and } \ C \in \mathcal{C}^{\hat{b}} \}.\]  Let $X^* = \{(q,t): t \leq \tau(q)\}$ denote the set of $(q,t)$ pairs that fall below the frontier defined by $\tau$. The delegate is allowed to combine any allowable contracts from $X^*$ into a menu, so the space of admissible menus is defined as $\mathcal{C}^* = \{ C \subseteq X^* \mid (0,0) \in C\}$. We depict this translation from $\mathcal{C}^{\hat{b}}$ to the associated frontier $\mathcal{C}^*$ in \figref{fig: procurement}. A menu of outcomes is represented by the delegate's promised mapping from quantity to transfer. The optimal contracting space from \Propref{prop:procurement} requires the delegate to choose one menu from among $\mathcal{C}^{\hat{b}}$, which is one curve on the left panel. The outcome-based restriction in the price-quantity frontier, however, allows the delegate to choose any mapping from quantity to transfer, so long as it lies below $\tau(q)$.

\begin{figure}[t]
    \centering
    \begin{subfigure}[t]{0.47\textwidth} 
        \centering
        \begin{tikzpicture}[xscale=3.8, yscale=3.8]

            \draw[thick, orange] 
            plot[domain=0:1, samples=20] (\x, {(\x)^2});
            \node[right] at (1,1) {$C_{\hat{b}}$};
            \draw[thick, orange] 
            plot[domain=0:1, samples=20] (\x, {0.75*(\x)^2});
            
            \draw[thick, orange] 
            plot[domain=0:1, samples=20] (\x, {0.5*(\x)^2});
            
            \draw[thick, orange] 
            plot[domain=0:1, samples=20] (\x, {0.25*(\x)^2});
            \draw[->] (0,0) -- (1.1,0) node[below] {$q$};
            \draw[->] (0,0) -- (0,1.1) node[left] {$t$};
        \end{tikzpicture}
        \caption{Menu-based restriction}
        \label{fig: procurement a}
    \end{subfigure}
    \hspace{0.03\textwidth}
    \begin{subfigure}[t]{0.47\textwidth}
        \centering
        \begin{tikzpicture}[xscale=3.8, yscale=3.8]
            
            \fill[orange!30] (0,0) plot[domain=0:1, samples=20] (\x, {(\x)^2}) node[right, black] {$\tau(q)$}
                -- (1,0) -- cycle;
            \draw[thick, orange] plot[domain=0:1, samples=20] (\x, {(\x)^2});
            \draw[->] (0,0) -- (1.1,0) node[below] {$q$};
            \draw[->] (0,0) -- (0,1.1) node[left] {$t$};
        \end{tikzpicture}
        \caption{Price-quantity frontier}
        \label{fig: procurement b}
    \end{subfigure}
    \caption{Delegated Procurement}
    \label{fig: procurement}
    \captionsetup*{justification=raggedright,singlelinecheck=false}
    \caption*{ \scriptsize Note: We depict a menu by its associated mapping from quantity to transfer. On the left panel are the menu-based contracting restrictions from \Propref{prop:procurement}. On the right is the associated price-quantity frontier.}
\end{figure}

Our next result shows the principal does not lose by delegating $\mathcal{C}^*$, instead of the more restrictive $\mathcal{C}^{\hat{b}}$, when there are many high-cost agents.

\begin{proposition}
\label{prop: pq-frontier}
Suppose the density of the agent's type, $g$, is increasing. Then, delegating $\mathcal{C}^{\hat{b}}$ is outcome-equivalent to delegating $\mathcal{C}^*$. That is, for each $b \in [0,1]$, the $b$-delegate's preferred menu from $\mathcal{C}^*$ belongs to $\mathcal{C}^{\hat{b}}$.
\end{proposition}
\medskip

For $b \leq \hat b$, the claim that all delegates choose the menu $C_b$ from $\mathcal{C}^*$ is immediate regardless of $g$, as $C_b$ was defined to be $b$-optimal. For types above $\hat{b}$, this result is less straightforward and, in general, not true. For an illustration of the problem arising when $b > \hat{b}$, consider a menu $C_{\hat{b}}$ such that the agent's quantity $q_{\hat{b}}(s)$ is decreasing in the agent's cost type. Suppose that the delegate offered the menu $C_{\hat{b}}$, but removed the quantities intended for types $s \in (s',s'')$ from the menu altogether. Low-cost agents in this interval would increase production to $q_{\hat{b}}(s')$ while higher-cost agents would reduce production to $q_{\hat{b}}(s'')$. If there were sufficiently many low-cost agents within this interval, the expected quantity procured under this new menu could be strictly higher than under $C_{\hat{b}}$, potentially making this deviation profitable for the highest-type delegates. Nevertheless, under the assumption that the density $g$ is increasing, there are relatively more high-cost sellers than low-cost sellers, making this type of deviation unprofitable for the delegate.


\subsection{Efficiency through Delegation} \label{sec: efficiency}

This section studies under what conditions efficient outcomes can be implemented via delegated contracting. We highlight fundamental differences in how players' outside options and private information constrain delegated contracting relative to centralized mechanisms, shedding light on the cost of delegation. These results speak to the interaction between contracting authority and property rights, which appears across a wide range of economic domains including industrial organization, antitrust regulation, and the boundaries of the firm.\footnote{We direct the reader to \citet{dworczak_mechanism-design_2024} for several applications and a thorough discussion of this literature.}

We focus on a standard budget-balanced allocation problem. The principal wishes to efficiently allocate one unit of a good between two players. Player $i \in \{1,2\}$ is privately informed about his willingness-to-pay for the good, $\theta_i \in [0,1]$, drawn independently according to a CDF $F_i$ with full support and density $f_i$. The outcome space consists of $(q_i,t_i)_{i \in \{1,2\}}$, where $q_i$ represents the fraction of the good to be ultimately allocated to player $i$, and $t_i$ is the transfer paid by player $i$. An outcome is feasible if at most one unit is allocated: $q_i\geq 0$, with $q_1+q_2\leq 1$; and transfers are balanced: $t_1+t_2 = 0$. Player $i$'s payoff when his type is $\theta_i$ and the outcome includes $(q_i, t_i)$ is $\theta_i q_i - t_i.$

Each player initially owns a share $r_i \geq 0$ of the good, with $r_1 + r_2 \leq 1$. Player $i$'s outside option when his type is $\theta_i$ is $r_i \theta_i$. The principal owns the remaining $1-r_1-r_2$ shares, and aims to implement the efficient ex-post allocation through delegated contracting. In our simple linear setting, this means allocating the good to the player who values it the most: $q_i = \mathbf{1}[\theta_i \geq \theta_{-i}]$, with ties broken arbitrarily if $\theta_1=\theta_2$.

By varying the ownership structure $(r_1,r_2)$, our model encompasses several canonical settings. We briefly review the primary benchmarks pertaining to centralized mechanisms before exploring delegated implementation.

\paragraph{Example 1 (Principal Ownership).} The principal initially owns the good to be allocated, that is, $r_1=r_2=0$. In a centralized mechanism, the efficient allocation can be obtained by running an efficient auction and appropriately distributing the proceeds to the losing party \citep{dAspremont1979}.\hfill $\blacksquare$
\paragraph{Example 2 (Partnership).} The two players jointly own an asset, characterized by ownership shares $(r_1,r_2)$, where $r_1+r_2=1$. 
The standard bilateral trade problem \citep{myerson_efficient_1983} corresponds to the extreme case in which $r_1 = 0$ and $r_2 = 1$. 
When the players share the same (non-degenerate) distribution of values $F_1 = F_2 = F$, no efficient and budget-balanced mechanism exists. 
On the other hand, \citet{cramton_dissolving_1987} show that when $r_1$ and $r_2$ are sufficiently close, the partnership can be efficiently dissolved. In particular, efficient and budget-balanced dissolution is possible for any value distribution $F$ whenever $\max\{r_1, r_2\} \leq \frac{3}{4}$ \citep{fieseler2003partnerships}. \hfill $\blacksquare$

\medskip

We now investigate efficient implementation. As in our original setup, we proceed under the assumption that Player 1 is the delegate, who offers a menu to Player 2, the downstream agent.

Applying \theoref{theo: implementability}, we show that whenever the efficient allocation is implementable through delegation, there is a unique contracting space that implements it.\footnote{By ``unique'', we mean there is only one set of outcomes $(q_i,t_i)_{i \in \{1,2\}}$ that  occur on path. Of course, one could always add dominated menus or outcomes that are never chosen on path without disturbing implementation. If the lowest type in the support is strictly positive, the contracting space need not be unique, but the following results in this section still hold.} We introduce two pieces of useful notation to describe this space. Let $\mu_i := \int v\ dF_i(v)$ denote player $i$'s average type and $I_{F_{i}} (x) := \int^x_0 F_{i}(v)\ dv$ denote the integral CDF of $F_i$ evaluated at the type $x$.

The unique contracting space described above is best understood as a collection of bid-ask pairs. Specifically, $\mathcal{C} = \{C^\lambda \}_{\lambda \in [0,1]}$, whose constituent menus are indexed by a parameter $\lambda \in [0,1]$. By choosing a menu $C^\lambda$, Player 1 commits to a takeover ``bid'' and a takeover ``ask'', $b^\lambda$ and $a^\lambda$, respectively. Player 2 has the right to forfeit all his shares and receive payment $b^\lambda$, or purchase all of Player 1's shares at price $a^\lambda$. Moreover, the takeover bid and ask amounts are specified by:
\begin{align*}
    b^\lambda = \lambda - I_{F_2} (\lambda), \qquad \text{ and } \qquad 
    a^\lambda = I_{F_2} (\lambda).
\end{align*}
\noindent This contracting space leads to the following ex-post payoffs:
\begin{align}
    U_1(\theta_1,\theta_2) &= I_{F_2}(\theta_1) \label{eq: efficient_del_payoff}\\ 
    U_2(\theta_1,\theta_2) &= \max\{\theta_1,\theta_2\} - I_{F_2}(\theta_1) \label{eq: efficient_agent_payoff}
\end{align}

Notably, Player 1 is fully insured against Player 2's type, while Player 2 is the residual claimant to the social surplus. The integral CDF term $I_{F_2}(\theta_1)$ arises from the standard application of Bayesian incentive compatibility, reflecting the interim expected surplus given Player 1's type, up to a constant. The characterization above only relies on the outside options at 0, which are independent of the true ownership shares $(r_1,r_2)$.  Therefore, the ownership shares that permit efficient implementation are those that make these payoffs individually rational.

\begin{proposition}\label{prop: efficient implementation general}
    Fix $F_1,F_2$. The efficient allocation can be implemented through delegation if and only if ownership shares $(r_1,r_2)$ satisfy $r_1 = 0$ and $r_2 \leq \mu_2.$    
\end{proposition}

The ownership shares that permit efficient outcomes are quite restrictive and place an asymmetric burden on Player 1 (delegate) and Player 2 (agent). While Player 1 cannot hold any fraction of the good, Player 2's share  must be smaller than the average of his value distribution.

If the delegate, Player 1, holds any share of the asset, his participation decision is violated for sufficiently low types $\theta_1$. In the mechanism, Player 1's payoff is given by $U_1(\theta_1) = I_{F_2}(\theta_1)$. Because $U_1'(0) = F_2(0) = 0$, Player 1's payoff is locally constant for the lowest types. On the other hand, his outside option grows linearly in proportion to his ownership share $r_1$. If $r_1 > 0$, then sufficiently low types anticipate expropriation and will not participate. These low types expect to relinquish the asset with high probability, but the takeover ask amount is too low to appropriately compensate them.

Unlike Player 1, Player 2 can be convinced to participate while owning a positive share of the asset, so long as that share is not too large. One intuition is that Player 2 is the residual claimant on the social surplus, which is positive even when his type is zero. In fact, his participation constraint is tightest when the types $\theta_1$ and $\theta_2$ are both large and close together. In this case, Player 2 may be asked to relinquish the good even though his value is arbitrarily close to 1, despite the takeover bid being below 1, since $b^1 = 1- I_{F_2}(1) =  \mu_2$. This imposes an upper bound on Player 2's shares.
Applying \Propref{prop: efficient implementation general} to the two canonical settings described above delineates the cost of delegation.
\begin{corollary} \label{cor: efficient implementation}
    Under \textbf{principal ownership}, efficiency is implementable through delegation for all type distributions.
    
    No \textbf{partnership} can be efficiently dissolved through delegation, regardless of the type distribution.     
\end{corollary}
Contrasting Corollary~\ref{cor: efficient implementation} with the centralized benchmarks shows that mediation provided by centralized mechanisms is indispensable in the partnership setting, but unnecessary under principal ownership. In particular, the result of \citet{cramton_dissolving_1987}, described above, relies on the existence of a benevolent mediator who not only sets the rules for the interaction between the two partners, but also runs the bidding game that implements efficiency. Corollary~\ref{cor: efficient implementation} shows that the strategic uncertainty facilitated by the mediator is critical. The designer can never achieve efficient dissolution if she cannot directly mediate the interaction between the parties, regardless of the distribution of ownership.
\paragraph*{Whom to delegate to?}
\Propref{prop: efficient implementation general} suggests that the allocation of contracting rights is central for efficient implementation. That is, the choice of \textit{which player should be the delegate} is an important design variable. This problem has received little attention beyond \citet{severinov2008}, who introduces it within a team-production environment where subcontracting cannot be restricted by the principal. Our setup additionally allows us to study the effects of tailoring contractual restrictions to the chosen delegate.

We focus on the partnership setting where efficient implementation is impossible under budget balance (\corref{cor: efficient implementation}). Requiring the contracting space to still implement the efficient allocation, we now relax budget balance.
We then ask who is \textbf{the optimal delegate}: that is, which player $i \in \{1,2\}$, when acting as the delegate, allows efficient implementation at the minimal budget deficit.\footnote{When studying the allocation of property rights, \citet{Segal2016Property} also follow the approach of minimizing subsidies to guarantee efficiency.} In what follows, $j \in \{d,a\}$ indexes which player acts as the delegate or the agent.

The role of the principal's subsidies is to guarantee participation for the two players because incentive compatibility still pins down transfers up to a constant. It will be convenient to separate the transfers paid to guarantee the delegate's participation, denoted by $\tau^d$, from those required for the agent, $\tau^a$.
Comparing the delegate's interim payoff from \eqref{eq: efficient_del_payoff} to their outside option identifies the minimal expected subsidy for the delegate:
$\tau^d = - \min_{\theta_d \in [0,1]} \{I_{F_{a}}(\theta_d)-r_d \theta_d\}    
$.

Because the agent's participation decision is ex-post, the subsidy required also depends on the delegate's type. Comparing the agent's ex-post payoff in \eqref{eq: efficient_agent_payoff} to their outside option pins down the minimal delegate-type-dependent subsidy.
\begin{align*} 
\begin{split}
        \tau^a (\theta_d) &= - \min_{\theta_{a} \in [0,1]} \big\{ \max\{\theta_a, \theta_d\} - I_{F_{a}}(\theta_d) - r_{a} \theta_{a}\big\} =  I_{F_{a}} (\theta_{d})-(1-r_{a})\theta_{d}. 
\end{split}
\end{align*}

Combining the observations above, Equation \eqref{eq: total subsidy} below characterizes the total expected subsidy required to implement the efficient allocation for a given delegate.
\begin{align} \label{eq: total subsidy}
    \mathbf{E}[\tau^d + \tau^a] = K + (1-r_{d})\mu_d - \min_{\theta_d}\{ I_{F_{a}}(\theta_d) - r_d\theta_d\}, 
\end{align}
\noindent where $K < 0$ is a constant that is independent of the identities of the delegate and agent.
Switching the players' roles between delegate and agent is now a matter of identifying whether the above expression is larger with indices $(d,a) = (1,2)$ or $(d,a)=(2,1)$. Our first result establishes the optimal delegate when players are symmetric in terms of ownership, but differ in terms of information.

\begin{proposition}\label{prop: whom to delegate to --- info}
    Suppose ownership is symmetric: $r_1=r_2 = \frac{1}{2}$. If $F_1$ is a mean-preserving spread of $F_2$, then Player 1 is the optimal delegate.
\end{proposition}

The intuition for this result builds upon that of \Propref{prop: efficient implementation general}. As the agent's type distribution becomes more dispersed, the delegate becomes more uncertain about whether he will ultimately hold the asset. The increased possibility that he might retain the asset even when his type is low relaxes the delegate's participation constraint, thereby reducing the required subsidy. On the other hand, because the agent's subsidy moves linearly with the delegate's type, only the mean of the delegate's distribution matters. Given that $F_1$ and $F_2$ are assumed to have the same mean, the agent's expected subsidy is unchanged.

Our second comparison studies the case where the players are equal in terms of information, but differ in ownership. In this case, the shape of the common type distribution plays a role in choosing the optimal delegate.

\begin{proposition} \label{prop: whom to delegate to --- ownership}
    Suppose $r_1<r_2$ and both players' values are drawn according to $F$ with density $f$. If $f$ is increasing (decreasing), then 1 (2) is the optimal delegate. 
    
\noindent Moreover, if the designer can jointly design the contract space and the distribution of initial ownership, the optimal ownership share of the delegate is $r=F^{-1}(\mu).$
\end{proposition}

The identity of the optimal delegate depends crucially on the shape of the distribution of values. Contracting rights may be optimally given to minority owners and, in contrast to standard partnership dissolution problems, the optimal distribution of ownership rights is often asymmetric. The intuition behind this result---and why the shape of the distribution matters---can be seen through Equation \eqref{eq: total subsidy}. In particular, Equation \eqref{eq: total subsidy} identifies a tradeoff that arises as the delegate's share $r_d$ increases: the agent's required subsidy decreases, while the delegate's subsidy increases. The rate of increase in the agent's subsidy depends on the mean of the type distribution, because the agent is the residual claimant to the social surplus. On the other hand, the delegate's participation constraint binds at the $r_d$-quantile of the distribution $F$, where his uncertainty over buying or selling the asset is commensurate to his initial ownership. By the envelope theorem, as the delegate's share $r_d$ increases, the required subsidy increases according to the $r_d$'th quantile, $F^{-1}(r_d)$. Which of these two effects dominates determines which player is the optimal delegate. Roughly, when the shared density $f$ is increasing, there is a long left tail, and the mean is small relative to the quantiles. Therefore, the benefit of relaxing the delegate's participation outweighs the cost of tightening the agent's, and contracting rights should be assigned to the lower-share owner.

Overall, these results emphasize the importance of jointly studying contracting rights, property rights, and private information. Partnership models are a workhorse in incomplete-information industrial organization \citep{LoertscherMarx2022incomplete, Loertscher2026bargaining}. To this literature, we contribute a regulatory perspective by highlighting what kinds of external constraints foster efficient outcomes. More directly, our results inform the design of buy-sell clauses, deadlock-resolution mechanisms sometimes included in partnership agreements \citep{MCAFEE1992Dissolving, loertscher_optimal_2019}. The ``Texas Shootout'' is a typical implementation whereby one party (the delegate) proposes a price, and the other (the agent) chooses whether to buy or sell at that price. This mechanism is inefficient.
Our construction shows that efficiency requires separating and constraining the buy and sell prices to mitigate the proposer's monopoly/monopsony power. In addition, our results describe which partner should have the proposal power.


\subsection{Delegated Contracts for Resale}

We now study a stylized model of market intermediation. A revenue-maximizing seller (the principal) wishes to place a product in the market, but must do so through an informed intermediary (the delegate) who ultimately resells the product to a final consumer (the agent). Both the final consumer and the intermediary value the good and have standard unit-demand, quasilinear preferences. If the intermediary retains the product, he might privately enjoy it, resell it later in a non-contractible manner, or scrap it for its ``salvage value''. The intermediary's private value when allocated the good is $\theta_1 \sim G \in \Delta[0,1]$, and the consumer's private value is $\theta_2 \sim F \in \Delta[0,1]$. We assume $\theta_1$ and $\theta_2$ are independent, that both distributions $F$ and $G$ admit full support densities, $f$ and $g$ respectively, and have increasing hazard rates. We denote the virtual valuation function for a distribution $H$ with density $h$ as $\psi_H(x) := x - \frac{1-H(x)}{h(x)}$. By the assumption of increasing hazard rates, $\psi_F$ and $\psi_G$ are strictly increasing.

As is often the case in intermediated resale settings, the downstream consumer  has stronger demand than the intermediary. We capture stochastic gains from trade by assuming the distribution of consumer valuations conditional on any price stochastically dominates that of the intermediary: $F( v \mid v\geq p) \leq G ( v \mid v \geq p)$, for all $p \in [0,1]$ and $v\geq p$.\footnote{This assumption is stronger than first order stochastic dominance, but weaker than monotone likelihood ratio dominance.} That is, for any price $p$, the consumers who are willing to pay that price have stronger valuations than the intermediaries willing to keep the product at that price. In particular, this implies $\psi_F(x) \geq \psi_G(x)$.

The timing of the interaction follows our general setup. The seller first contracts with the intermediary, deciding whether to sell the product to the intermediary and constraining how it can be resold to the final consumer. The outcome space simply consists of an allocation of the good to at most one player and a net transfer paid by each player. Formally, an outcome is a tuple $(q_1, q_2, t_1, t_2)$, satisfying $q_1+q_2 \leq 1$, in which $q_i \in \{0,1\}$ represents whether player $i$ ends up with the good, and $t_i$ is the net transfer paid by that player. When $q_1=q_2=0$, the principal retains the good.

Before turning to the seller's optimal contract, we start by discussing the standard \textit{laissez-faire} benchmark. In this benchmark, the contract between the seller and the intermediary cannot specify a resale policy, and the seller's problem is solved by backward induction. The intermediary with type $\theta_1$ is a monopolist over the downstream consumer and optimally posts the monopoly price $p_m(\theta_1)= \psi^{-1}_F \left(\theta_1\right).$  The intermediary's resulting profits, inclusive of his value when retaining the good, are denoted by $\pi_m (\theta_1)$.
From the point-of-view of the seller, she effectively sells a good to a population of intermediaries with valuation $v = \pi_m(\theta_1)$ distributed according to $H= G \circ \pi^{-1}_m$. 
Thus, the seller maximizes revenue by selling to the intermediary at a \textit{laissez-faire} posted price $p_{LF}$ that is optimal against the distribution $H$. \figref{fig:sale-resale}, panel \subref{fig: sale-resale a} illustrates this allocation when all types are uniformly distributed. The figure shows the sets of $(\theta_1,\theta_2)$ under which the seller (the S cell, in gray), intermediary (the I cell, in orange) or consumer (the C cell, in blue) hold the good. 
\begin{figure}[t]
\centering

\begin{subfigure}[t]{0.47\textwidth} 
\centering

\begin{tikzpicture}[xscale=3.8, yscale=3.8]
    \draw[gray] (0,0) rectangle (1,1);

    \draw[thick, <->] (0,1.1) -- (0,0) -- (1.1,0);
    \node[left] at (0,1.12) {\small $\theta_2$};
    \node[below] at (1.15,0) {\small $\theta_1$};

    \node[left] at (0,1) {\small $1$};
    \node[below] at (1,0) {\small $1$};

    \draw[fill=first_best, opacity=.3] 
        (.33,.5*1.33) -- (1,1) -- (.33,1) -- cycle;

    \draw[fill=orange, opacity=.3] 
        (.33,0) -- (.33,.5*1.33) -- (1,1) -- (1,0) -- cycle;

    \draw[
        pattern={Lines[angle=135, line width=0.2pt, distance=3pt]},
        pattern color=black
    ] 
        (.33,.33) -- (.33,.5*1.33) -- (1,1) -- cycle;

    \draw[fill=cream, opacity=.3] 
        (0,0) -- (.33,0) -- (.33,1) -- (0,1) -- cycle;

    \draw[
        pattern={Lines[angle=135, line width=0.2pt, distance=3pt]},
        pattern color=red
    ] 
        (0,0) -- (.33,0) -- (.33,1) -- (0,1) -- cycle;

\node at (.55,.9) {C};
\node at (.72,.4) {I};
\node at (.16,.50) {S};

    \draw[thin, dashed, gray] (0,0) -- (1,1);
    \draw[gray] (.33,0) -- (.33,1);
\end{tikzpicture}
\caption{Laissez-faire allocation}
\label{fig: sale-resale a}
\end{subfigure}
\hspace{0.03\textwidth}
\begin{subfigure}[t]{0.47\textwidth}
\centering
\begin{tikzpicture}[xscale=3.8, yscale=3.8]
    \draw[gray] (0,0) rectangle (1,1);

    \draw[thick, <->] (0,1.1) -- (0,0) -- (1.1,0);
    \node[left] at (0,1.12) {\small $\theta_2$};
    \node[below] at (1.15,0) {\small $\theta_1$};

    \node[left] at (0,1) {\small $1$};
    \node[below] at (1,0) {\small $1$};
    \node[left] at (0,.5) {\small $\frac{1}{2}$};
    \node[below] at (.5,0) {\small $\frac{1}{2}$};

    \draw[fill=first_best, opacity=.3] 
        (0,.5) -- (.5,.5) -- (1,1) -- (0,1) -- cycle;

    \draw[fill=orange, opacity=.3] 
        (.5,0) -- (.5,.5) -- (1,1) -- (1,0) -- cycle;

    \draw[fill=cream, opacity=.3] 
        (0,0) -- (.5,0) -- (.5,.5) -- (0,.5) -- cycle;

    \draw[thin, dashed, gray] (0,0) -- (.5,.5);

    \draw[thick, red, domain=.5:1, samples=100] 
        plot (\x, {\x -.5*(\x)^2 + .125});

    \draw[thin, red] (0,.5) -- (.5,.5);

    \node at (.35,.77) {C};
\node at (.75,.35) {I};
\node at (.25,.25) {S};
\end{tikzpicture}
\caption{Optimal allocation}
\label{fig: sale-resale b}
\end{subfigure}

\caption{Delegated contracts for resale.}
\label{fig:sale-resale}
\captionsetup*{justification=raggedright,singlelinecheck=false}
\caption*{ \scriptsize Note: The picture illustrates allocations when the intermediary's and the consumer's types are distributed uniformly on $[0,1]$. The shaded cells indicate the final owner of the good: S for seller, I for intermediary, and C for consumer. The left panel depicts the laissez-faire allocation, with hatched areas highlighting inefficiencies. The right panel depicts the optimal allocation as in \Propref{prop: resale with buyback}, with the red line plotting the wholesale price chosen by each type of the intermediary.}
\end{figure}

Generically, \textit{laissez-faire} contracting leads to two inefficiencies. First, the intermediary leverages his market power to the extreme, acting as a full monopolist and retaining the good even when the valuation of the final consumer exceeds his own. This is an intensive margin distortion, represented in \figref{fig:sale-resale}-\subref{fig: sale-resale a} as the area hatched in black between the I and C cells. Second, the seller typically posts a price that excludes intermediaries with a lower valuation from the market. This is an extensive margin distortion, because some intermediaries do not trade, and is represented by the area hatched in red on the S cell. Because this extensive margin distortion breaks the intermediation chain, the good may fail to reach high willingness-to-pay consumers. We will show that the optimal delegated contract space  partially addresses both kinds of inefficiencies.

To determine how the seller should optimally constrain the intermediary's resale, we appeal to \theoref{theo: implementability}. This recasts our problem as an auctioneer facing two heterogeneous consumers, allowing us to immediately invoke standard results. In this case, the revenue-maximizing auction allocates the good to the agent with the highest virtual value, when positive; otherwise, the seller retains the good. This asymmetric auction maximizes the seller's revenue under BIC and IIR constraints but, importantly, can also be made DSIC and EPIR for both bidders. \theoref{theo: implementability} implies that the same revenue can be achieved by correctly designing the intermediary's contracting space.  Although the physical description of our environment---in which the intermediary holds the good before the consumer---suggests that the seller must cede excess rents to the intermediary, the optimal auction allocation shows this need not be the case. 


We describe an indirect contract space that implements the optimal asymmetric auction as an equilibrium. Each allowable contract can be thought of as a combination of two types of clauses: price agreements and buyback policies. A \textbf{price agreement} with resale price $p$ is a procedure whereby the intermediary agrees to resell the good at price $p$ after buying the good from the seller at discounted price $p - d(p)$. A \textbf{buyback policy} with refund $r$ is a commitment from the seller to rebuy any unsold unit at price $r$. A menu implementing a price agreement and buyback policy takes the following form:
$$C = \left\{\underbrace{(1,0,p-d(p),0),}_{\text{intermediary keeps the good}} \underbrace{(0,1,-d(p),p),}_{\text{buyer purchases}}  \ \underbrace{(0,0,p-d(p)-r,0)}_{\text{seller buys back}} \right\}.$$

\smallskip

\begin{proposition} \label{prop: resale with buyback}
    The seller's expected revenue is maximized by a contracting space $\mathcal{C} = \{C^p\}_{p \in [\dw p,1]}$ with $\dw p = \psi^{-1}_F (0)$. Each menu $C^p$ is a price agreement with buyback policy $r$, such that for resale price $p$:
    
    \begin{enumerate}
    \item The discount $d(p) \in [0,p]$ is given by

$$d(p) := p F(p) - r F(\dw p) - \int^p_{\dw p} \psi^{-1}_G \circ \psi_F (x) f(x) dx.  $$

 \item The buyback policy's refund is $r := \psi^{-1}_G (0)$.
    
    \end{enumerate}

\end{proposition}

The optimal delegation of resale limits the two inefficiencies that arise in a \textit{laissez-faire} setting. First, it reduces the ability of the intermediary to exercise market power on the intensive margin. Indeed, intermediaries optimally choose resale prices below the monopoly price: $p(\theta_1) = \psi^{-1}_F \left( \psi_G(\theta_1) \right) \leq \psi^{-1}_F (\theta_1) = p^m (\theta_1)$. Second, it never excludes high-value downstream consumers. These consumers may not receive the good under \textit{laissez-faire} because low-value intermediaries, wary of being unable to resell the good, do not purchase in the first place. The seller circumvents this problem through the buyback policy. When the intermediary has low value, he buys the good knowing he can return it at zero expected profits. In this way, the seller can replicate the optimal auction through delegated contracting. \figref{fig:sale-resale}-\subref{fig: sale-resale b} plots the allocation under the optimal contracting space with uniform private information. Because types are identically distributed, there is no intensive margin distortion. The red line reflects the wholesale price an intermediary with type $\theta_1$ pays to the seller, and the buyback policy has $r=1/2$.

Intuitively, there are two potential sources of loss for the seller in this resale environment: private information and monopoly power of the intermediary. Under \textit{laissez-faire}, these two sources compound. However, under delegated contracting, the seller can neutralize the intermediary's monopoly power. What remains are rents ceded to private information, which are unavoidable.

\section{Conclusion}
This paper takes a first step toward a unified framework for studying delegated contracting problems. \theoref{theo: implementability} and its extensions offer guidance on the design of optimal restrictions in a large set of applications involving vertical controls in organizations and regulation. While the applications in this paper  illustrate both the scope of the framework and the gap between centralized and delegated implementation, many natural questions remain. What can the principal achieve when menu-based restrictions are unavailable, but outcome-based constraints can be applied? How can the designer mitigate concerns about collusion between the delegate and the agent? And how should regulation adapt to a sequence of downstream interactions? 

\appendix
\section{Proofs} \label{sec: proofs}
\subsection*{Proof of \theoref{theo: implementability}}
\label{proof:theo:implementability}

\subsubsection*{Only if  $\Rightarrow$} Assume that $Y$ is implementable through delegation. Let $\mathcal{C}$ be the contract space, and $\left(\sigma_1, \{\sigma_2^C,\mu^C\}_{C \in \mathcal{C}}\right)$ an associated equilibrium that implements $Y$. Because any equilibrium is a Bayes-Nash Equilibrium, the revelation principle implies that there exists a truthful direct centralized mechanism which Bayes-implements $Y$. Formally, there exists $\mathcal{M} = \left(M, g \right)$ such that $M = \Theta_1 \times \Theta_2$, with optimal strategies $s_i (\theta_i) = \theta_i$, for $i \in \{1, 2\}$. Because this equilibrium implements $Y$, it must be the case that $g(\theta_1,\theta_2) = Y(\theta_1,\theta_2)$ for all $(\theta_1,\theta_2) \in \supp \mu_o$. Moreover, because $Y$ is deterministic, $\sigma^C_2(\theta_2)$ is deterministic for all $C \in \supp \sigma_1(\theta_1)$ and because $Y$ is implemented through delegation:
\begin{equation} \label{eq: outcome transform}
 g (\theta_1, \theta_2) = \sigma^C_2(\theta_2), 
\end{equation}
for all $(\theta_1,\theta_2) \in \supp \mu_o$, $C \in \supp \sigma_1 (\theta_1)$.

Because $Y$ is implemented by $\mathcal{M}$, it must satisfy BIC and IIR for both players. In what follows we prove it also satisfies (i) DSIC and (ii) EPIR for the agent.

Fix any $\theta_1 \in \Theta_1$, and $C \in \supp \sigma_1(\theta_1)$. Because $\left(\sigma_1, \{\sigma_2^C,\mu^C\}_{C \in \mathcal{C}}\right)$ was an equilibrium in the delegated contracting game, we have, for any agent's information $\theta_2$ such that $(\theta_1,\theta_2) \in \supp \mu_o$ and any alternative outcome available to the agent, $x' \in C$:
\begin{align*}
\begin{split}
u_2 (\sigma^C_2(\theta_2), \theta_2) = \int_{\hat\theta_1} u_2 \left( \sigma^C_2 (\theta_2), \theta_2  \right) \mu^C(d\hat\theta_1|\theta_2) \\   \geq \int_{\hat\theta_1} u_2 \left( x', \theta_2  \right) \mu^C(d\hat\theta_1|\theta_2)  = u_2 (x', \theta_2), 
\end{split}
\end{align*}
where the first and last equalities follow from $\theta_1$ being payoff-irrelevant for player 2. We can, in particular, choose $x' = \sigma^C_2 (\theta'_2)$ for any type of the agent with $(\theta_1,\theta'_2) \in \supp \mu_o$, and take expectation over $C$ on both sides with respect to $\sigma_1 (\theta_1)$ to obtain:
\begin{align*}
\begin{split}
u_2 \left(g(\theta_1,\theta_2),\theta_2 \right)     = \int_C  u_2 \left( \sigma^C_2(\theta_2), \theta_2  \right)   \sigma_1(\theta_1) (dC) \\  \geq  \int_C  u_2 \left( \sigma^C_2(\theta'_2), \theta_2  \right)   \sigma_1(\theta_1) (dC)  = u_2 (g(\theta_1, \theta'_2), \theta_2),  
\end{split}
\end{align*}
where the first and last equalities come from \ref{eq: outcome transform}. Because $\theta_1$ was arbitrary, the inequality above shows truthful revelation is a dominant strategy for the agent in direct mechanism $\mathcal{M}$. 

Because we assume the agent can always choose $o$, the result above also implies, for any $\theta_1$:
$$u_2 (g(\theta_1,\theta_2),\theta_2)  \geq u_2 (o, \theta_2),$$
so that truthful revelation also satisfies ex-post individual rationality. Because, by Bayes implementation, $g=Y$, we obtain that $Y$ satisfies $(i)$ and $(ii)$. 

\subsubsection*{If $\Leftarrow$} To prove the converse, let $Y$ be such that $(i)$ and $(ii)$ hold. By the revelation principle, the $Y$-associated direct mechanism $\mathcal{M} = (M,g)$ implements $Y$ in a truthful-revealing equilibrium. Because it implements $Y$, we must have:
\begin{equation} \label{eq: outcome transform 2}
g(\theta_1,\theta_2) = Y(\theta_1,\theta_2).    
\end{equation}
Consider the contract space $\mathcal{C} = (C^{\theta_1})_{\theta_1 \in \Theta_1}$, where $$C^{\theta_1} = \{y: \exists \theta_2 \text{ with } (\theta_1,\theta_2) \in \supp \mu_o \text{ and } y=Y(\theta_1,\theta_2)\}.$$ 
Given this contract space, we will show that there is a PBE for the delegation game for each type of the delegate to offer menu $C^{\theta_1}$, and, upon observing $C^{\theta_1}$, for the agent to choose contract $Y(\theta_1,\theta_2) \in C^{\theta_1}$. 

Starting with the agent's best response, because of DSIC, and because $\mathcal{M}$ implements $Y$, for all $\theta_1 \in \Theta_1$, $\theta_2, \theta'_2 \in \Theta_2$:
\begin{align*}
u_2(Y(\theta_1,\theta_2),\theta_2) = u_2(g(\theta_1,\theta_2),\theta_2) \\ \geq  u_2(g(\theta_1,\theta'_2),\theta_2)   = u_2(Y(\theta_1,\theta'_2),\theta_2) ,  
\end{align*}
where the equalities follow from equation \ref{eq: outcome transform 2}. 

Because of EPIR, and because $\mathcal{M}$ implements $Y$ for all $\theta_1 \in \Theta_1$, $\theta_2, \theta'_2 \in \Theta_2$:
\begin{align*} u_2(Y(\theta_1,\theta_2),\theta_2) = u_2(g(\theta_1,\theta_2), \theta_2) \geq  u_2(o,\theta_2)  ,  
\end{align*}

so the two inequalities above show that it is optimal for the agent to reveal his type truthfully in the delegation game for any contract $C^{\theta_1}$, regardless of the beliefs the agent might hold about the delegate. 

Next, because $\mathcal{M}$ Bayes-implements $Y$ for the delegate:
\begin{align*}
    \begin{split}
   \int_{\hat\theta_2} u_1(Y(\theta_1,\hat\theta_2),\theta_1)  \mu_o (\theta_1,d\hat\theta_2)   = \int_{\hat\theta_2}  u_1(g(\theta_1,\hat\theta_2),\theta_1)  \mu_o (\theta_1, d\hat\theta_2) \\ \geq        \int_{\hat\theta_2}  u_1(g(\theta_1',\hat\theta_2),\theta_1)  \mu_o (\theta_1, d\hat\theta_2) = \int_{\hat\theta_2}u_1(Y(\theta_1',\hat\theta_2),\theta_1)  \mu_o (\theta_1,d\hat\theta_2),
    \end{split}
\end{align*}
Moreover, by IIR: 
\begin{align*}
    \begin{split}
   \int_{\hat\theta_2} u_1(Y(\theta_1,\hat\theta_2),\theta_1)  \mu_o (\theta_1,d\hat\theta_2)   = \int_{\hat\theta_2}  u_1(g(\theta_1,\hat\theta_2),\theta_1)  \mu_o (\theta_1, d\hat\theta_2)\\ \geq  u_1(o,\theta_1),
    \end{split}
\end{align*}
implying it is an equilibrium of the delegated game for the delegate to choose the contract associated to his type, as long as the agent truthfully reveals his type. Choose beliefs $\mu^C$ to follow from Bayes' rule whenever possible, and to be arbitrary off-path. Because the agent is truthful under any belief he may hold, he is truthful for this particular belief. Moreover, the outcome implemented in this equilibrium is $Y (\theta_1,\theta_2)$, concluding the proof. \hfill $\blacksquare$

\subsection*{Proof of Proposition \ref{prop:procurement}}
\label{apx:procurement}
The structure of the proof is to start from the program given by \theoref{theo: implementability} and then replace the agent's incentive constraints with an upper bound on the delegate's utility. In the relaxed problem, we show that this bound will be binding for an interval of types $[0,\hat{b}]$ and slack thereafter. The value of this relaxation can be achieved by allowing all types to choose an optimal screening contract from $(C_{b}){_b \in [0,\hat{b}]}$. 

Using \theoref{theo: implementability}, the principal's full problem is the following.
\begin{align*}
    \max_{q,t} &\mathbb{E}_{b,s}[ v(b)q(b,s) - t(b,s)]\\
    s.t. \quad & \mathbb{E}_s[ b q(b,s) - t(b,s)] \geq \mathbb{E}_s[ b q(b',s) - t(b',s)] \quad \forall b,b' \tag{Delegate BIC}\\
    & \mathbb{E}_s[ b q(b,s) - t(b,s)] \geq 0 \quad \forall b \tag{Delegate IIR}\\
    & t(b,s) - c(s,q(b,s)) \geq t(b,s') - c(s,q(b,s')) \quad \forall b, s, s' \tag{Agent DSIC}\\
    & t(b,s) - c(s,q(b,s)) \geq 0 \tag{Agent EPIR}
\end{align*}
We move towards a relaxation using the solution to type-$b$ delegate's unconstrained screening problem. The unconstrained screening problem can be written:
\begin{align*}
    \max_{q(b,\cdot),t(b,\cdot)} &\mathbb{E}_{s}[ b q(b,s) - t(b,s)]\\
    s.t. \quad & t(b,s) - c(s,q(b,s)) \geq t(b,s') - c(s,q(b,s')) \quad \forall b, s, s' \tag{Agent DSIC}\\
    & t(b,s) - c(s,q(b,s)) \geq 0 \tag{Agent EPIR}
\end{align*}
Let $U^*(b)$ denote the value of the above problem for each $b$, let $Q^*(b) = \mathbb{E}_s[q^*(b,s)]$ denote the expected output under the $b$-optimal mechanism, and $T^*(b)$ the corresponding expected transfer. Note that each delegate faces the same set of incentive constraints from the downstream agent. Single crossing then implies $Q^*(b)$ is weakly increasing in $b$, and $U^*(b)$ is weakly increasing and convex, with $U^*(0) = 0$. 

Since the principal can only restrict the contract space, DSIC and EPIR of the agent imply that the delegate can never achieve a payoff above $U^*(b)$. Hence, we consider the following relaxed problem, where the principal solves for \textit{expected} quantity $Q(b)$ and expected transfer $T(b)$ associated with the type-$b$ delegate:
\begin{align*}
    \max_{Q(b),T(b)} &\mathbb{E}_{b}[ v(b) Q(b) - T(b)]\\
    s.t. \quad & \text{Delegate BIC, IIR}\\
    & U(b) \leq U^*(b) \quad \forall b
\end{align*}
We now proceed with the usual Myersonian transformation, as incentive compatibility in this relaxed problem is characterized by the envelope and monotonicity conditions. Standard substitution of the envelope condition yields the following program. 
\begin{align*}
\max_{Q(b)} \quad & \int_0^1 \phi(b) f(b) Q(b) \, db \\
\text{s.t.} \quad & Q(b) \geq 0 \quad \text{for all } b \in [0,1] \\
& Q(b) \text{ non-decreasing on } [0,1] \\
& \int_0^b Q(r)\, dr \leq \int_0^b Q^*(r)\, dr \quad \text{for all } b \in [0,1]
\end{align*}
Here, $\phi(b) := \frac{1-F(b)}{f(b)} - \left( b - v(b) \right)$ represents the virtual value of the principal. The function $Q^*(b)$ is the non-negative, non-decreasing expected quantity derived from the delegate's unconstrained screening problem.

We define the integrand $J(b) := \phi(b) f(b) = {1 -F(b)} - (b - v(b))f(b)$. We now establish the structure of the optimal solution in four steps.

\paragraph{Step 1: $J(b)$ crosses zero from positive to negative}
We first show that $J(b)$ is positive at $b=0$ and negative at $b=1$. At $b=0$, since $F(0)=0$ and $v(0) =  0$, we have $J(0)=1+v(0)f(0)>0.$ 
At $b=1$, since $F(1)=1$, we have $J(1)=-(1-v(1))f(1)<0$, which follows from the fact that $v(1)<1$, which must hold given $b > v(b)$ on a positive measure of $b$ and $b-v(b)$ is non-decreasing.

Since $J(b) = \phi(b) f(b)$ and $f > 0$, $J$ and $\phi$ have the same sign. Given the monotone hazard rate assumption and the assumption $b - v(b)$ is weakly increasing, $\phi$ is weakly decreasing in $b$. Hence, $\phi$ crosses 0 from positive to negative. 

Let $\bar{b} = \inf_b \{b : J(b) < 0\}$ denote the value of $b$ after which $J$ becomes negative. By continuity of $f$, we must have $\bar{b} < 1$.

\paragraph{Step 2: Optimally, $Q(b)$ is constant on $[\bar{b}, 1]$}

Take any feasible $Q$ for which there exist $b_1, b_2 \in [\up{b},1)$, with $Q(b_1)< Q(b_2)$. By monotonicity of $Q$, $b_2>b_1$ and $Q(b_1) < Q(b)$ for all $b\geq b_2$. Define a new function $\tilde{Q}$ that is identical to $Q$ on $[0, b_1]$ and constant thereafter:
\[
\tilde{Q}(b) = 
\begin{cases}
Q(b) & \text{if } b \leq b_1 \\
Q(b_1) & \text{if } b > b_1
\end{cases}
\]
Then $\tilde{Q}$ is non-decreasing, satisfies the same cumulative constraint (as $\tilde{Q}(b) \leq Q(b)$ for all $b$), and yields a strictly higher objective value because $J(b) < 0$ on $[b_2, 1]$ and $\tilde{Q}(b) < Q(b)$ there. Therefore $Q$ cannot be optimal. We conclude that any optimal $Q$ must be constant on $[\bar{b}, 1]$.

\paragraph{Step 3: $J(b)$ is decreasing on $\{ b : J(b) > 0 \}$}

We verify this directly using the expression $J(b) = {1 -F(b)} - (b - v(b))f(b) = \phi(b) f(b)$. Differentiating, we get:
\[
J'(b) = -f(b) - (1-v'(b))f(b) - (b-v(b))f'(b) = \phi'(b)f(b) + \phi(b)f'(b),
\]
We show that the equation above implies that $J'(b) < 0$ in two cases. If $f'(b) > 0$, then all terms in the first expression are negative because $b > v(b)$. If $f'(b) \leq 0$, then both terms in the second expression are negative because $\phi(b)>0$ and $\phi'(b)\leq0$ as in Step 1. Hence, $J(b)$ is decreasing wherever it is positive.

\paragraph{Step 4: The optimal $Q$ matches $Q^*$ up to a cutoff $\hat{b} \leq \bar{b}$}

Since $J(b)$ is decreasing on the relevant region $[0,\bar{b}]$, the marginal value of increasing $Q(b)$ is highest at the lowest types. Given that $Q^*$ is non-decreasing and the cumulative constraint is:
\[
\int_0^b Q(r) \, dr \leq \int_0^b Q^*(r) \, dr,
\]
the optimal solution must match $Q^*$ starting from $b = 0$ and continue doing so until the constraint binds. This follows from a standard monotone reallocation argument as in \citet{kleiner_extreme_2021}: when the integrand is decreasing, the optimum is to match the upper bound allocation as fully as possible starting from the left. 

Formally, the principal's problem can be decomposed into a two-step maximization problem: first deciding the maximal point at which $Q$ is no longer increasing, then deciding the level of $Q$ below this cutoff:
\begin{align*}
\max_{Q(b), \hat{b}} \quad & \int_0^{\hat{b}} J(b) Q(b) \, db + \int_{\hat{b}}^1 J(b) Q(\hat{b}) \, db\\
\text{s.t.} \quad & Q(b) \geq 0 \quad \text{for all } b \in [0,1] \\
& Q(b) \text{ non-decreasing on } [0,\hat{b}] \\
& \int_0^b Q(r)\, dr \leq \int_0^b Q^*(r)\, dr \quad \text{for all } b \in [0,\hat{b}]\\
&\int_0^{\hat{b}} Q(r)\, dr = \int_0^{\hat{b}} Q^*(r)\, dr
\end{align*}
For any given $\hat{b}$, the optimal $Q(b)$ follows from Proposition 3 in \citet{kleiner_extreme_2021}. Therefore, there exists a cutoff $\hat{b} \leq \bar{b}$ such that $Q(b) = Q^*(b)$ for all $b \leq \hat{b}$, and the cumulative constraint binds at $\hat{b}$:
\[
\int_0^{\hat{b}} Q(r)\, dr = \int_0^{\hat{b}} Q^*(r)\, dr.
\]
Thereafter, $Q(b)$ remains constant at $Q^*(\hat{b})$ for all $b > \hat{b}$.

Finally, this payoff can be implemented in the original problem by allowing each type of delegate to choose any optimal screening contract up to the level $\hat{b}$, at which point the quantity no longer changes. Choosing an optimal screening contract corresponds to choosing some pair of $(Q^*, T^*)$ that would occur in the delegate's unconstrained screening problem. In other words, each type of delegate solves \[\max_{\beta \leq \hat{b}} b Q^*(\beta) - T^*(\beta).\] All types $b\leq \hat{b}$ find it optimal to procure $Q^*(b)$, as this is their first-best contract. In this region, the integral constraint is binding. By single crossing, types above $\hat{b}$ find it optimal to procure $Q^*(\hat{b})$.  \hfill $\blacksquare$

\subsection*{Proof of Proposition \ref{prop: pq-frontier}}
\label{apx:procurement_pq-frontier}
The proof structure proceeds by showing that it is impossible for the delegate to procure more than $Q^*(\hat{b})$ units in expectation from any menu in $\mathcal{C}^*$. We study a relaxed direct-mechanism problem, where the agent is constrained to the maximal payoff she would earn under any menu in $\mathcal{C}^*$. When the density $g(s)$ is increasing, the solution to this problem coincides with the agent's payoffs under $C_{\hat{b}}$. Because $Q^*(\hat{b})$ is the maximal expected quantity that can be procured from any menu in $\mathcal{C}^*$, we conclude that all types above $\hat{b}$ find it optimal to offer $C_{\hat{b}}$. 

In what follows, we let $V(s;b)$ denote the agent's ex-post payoff when his type is $s$ and the delegate's type is $b$ under an arbitrary menu.
In our specification of the agent's problem, agent DSIC is equivalent to a non-increasing quantity rule $q(b,\cdot)$ and the envelope condition:
\begin{align*}
    V(s;b) &= V(1;b) + \int_s^1 c(q(b,r)) dr = t(b,s) - s c(q(b,s))
\end{align*}

For $b \leq \hat{b}$, we let $V^*(s;b)$ denote the agent's payoff when the delegate offers the $b$-optimal menu $C_b$.

\paragraph{Step 1: For all $b \leq \hat{b}$ and all $s$, $q^*(b,s)$ and $V^*(s ; b)$ are pointwise increasing in $b$.} 
Using the usual virtual-value representation of the delegate's problem, the unconstrained delegate's problem is:
\begin{align*}
    \max_{q(b,\cdot)} \int \bigg( b q(b,s) - \phi(s) c(q(b,s)) \bigg) g(s)\\
    s.t. \quad q(b,\cdot)\geq0 \ \ \text{and non-increasing}
\end{align*}
Note that the choice set of non-negative, non-increasing functions forms a lattice, and that the objective function features increasing differences in $(b,q)$. By Topkis Theorem (1978), the set of solutions is increasing in $b$ (in the strong set order) for every $s$. Let $q^*(b,\cdot)$ denote the pointwise maximal such solution for each $b$.
Given that $q^*(b,s)$ is weakly increasing in $b$, the envelope representation of the agent's payoff also implies that $V^*(s;b)$ is also weakly increasing in $b$ for every $s$.

\paragraph{Step 2: For any menu in $\mathcal{C}^*$, the type-$s$ agent's ex-post payoff is bounded above by $V^*(s; \hat{b})$.} Fix a $b$. Since $\left(q^*(b,s), t^*(b,s)\right)$ was optimal for the type-$s$ agent from $C_b$, the agent's payoff from any $(q,t) \in C_b$ is bounded above by $V^*(s;b)$. Since  $V^*(s;b)$ is increasing in $b$, the agent's payoff is then bounded above by $V^*(s;\hat{b})$. Therefore, the $s$-agent's preferred contract across all menus $(C_b)_{b \in [0, \hat{b}]}$ is her choice from $C_{\hat{b}}$. 

Finally, the definition of the price-quantity frontier $\tau$ implies that no other menu from $\mathcal{C}^*$ can give the agent a greater payoff. This is because each allowable contract in $X^*$ pays the agent no more than what she could have earned from some contract belonging to $\cup_{b \in [0,\hat{b}]} C_b$. 

\paragraph{Step 3: If $g(s)$ is increasing, then the expected quantity procured under any menu in $\mathcal{C}^*$ is bounded above by $Q^*(\hat{b})$} 
To put an upper bound on the expected quantity any type of delegate can procure by using any menu from $\mathcal{C}^*$, consider the following relaxed problem. 
\begin{align*}
    \max_{q(b,s)} &\int q(b,s) g(s) ds\\
    s.t. &\quad q\geq0 \ \text{ non-increasing}\\
    V(s;b) &= \int_s^1 c(q(b,r)) dr\\
    V(s;b) &\leq V^*(s;\hat{b}) \quad \forall s
\end{align*}
Note that the delegate's true type $b$ plays no role as we are not maximizing the delegate's payoff. Only the bound imposed by the $\hat{b}$-optimal menu enters.
We therefore omit the delegate's type $b$ in the sequel. To further ease notation, let $u := c^{-1}$. We now rewrite the upper bound on the agent's utility using the envelope characterization, 
%
and perform a change of variables such that $y(s) = c(q(s))$, so we are effectively choosing the agent's cost directly.  Substitutng, we obtain the following maximization 
\begin{align*}
    \max_{y(s)} &\int u\left(y(s) \right) g(s) ds\\
    s.t. \quad y&\geq0 \ \text{ non-increasing}\\
    \int_{{s}}^1y(r) dr &\leq \int_{{s}}^1 y^*(r) dr \quad \forall s,
\end{align*}
\noindent where $y^*(s) := c(q^*(\hat{b},s))$ denotes the agent's cost under the menu $C_{\hat{b}}$.

We argue that the solution to this problem requires $y(s) = y^*(s)$ for all $s$; that is, the agent's upper bound utility constraint is everywhere binding. The intuition is straightforward. Since $c$ is convex, $u$ is concave, meaning it is increasing the fastest for low values of $y$, which are associated with higher types $s$. Additionally, since $g$ is increasing, the benefit of increasing $y$ is larger for higher values of $s$. Combining the two, it is optimal for the constraint to bind for all $s$.

Towards a formal proof, first note that at any solution, $\int_{\underline{s}}^1 y(r) dr = \int_{\underline{s}}^1 y^*(r)dr$. If this inequality is slack on an interval around $\underline{s}$, then raising $y$ in this interval by a constant improves the objective without violating the monotonicity constraint. Let $Y^* := \int_{\underline{s}}^1 y^*(r)dr > 0$. We now show that for any feasible $y$,
\[  \int_{\underline{s}}^1 \left[ u(y^*(s)) - u(y(s)) \right] g(s)  \geq 0\]  
First, concavity of $u$ implies:
\[\int_{\underline{s}}^1 \big[ u(y^*(s)) - u(y(s)) \big] g(s) \ ds\geq \int_{\underline{s}}^1 u'(y^*(s)) \big[ y^*(s) - y(s) \big] g(s) \ ds \]
Next, let $w(s) = u'(y^*(s)) g(s) > 0$. It is sufficient to show that 
\begin{equation}
\int_{\underline{s}}^1 w(s) \frac{y^*(s) - y(s)}{Y^*} \geq 0 \label{eq: prop_FOSD}    
\end{equation}
The term $w(s)$ is increasing in $s$. This follows because $y^*$ is decreasing and $u$ is concave, so $u'(y^*(s))$ is non-decreasing in $s$. Since $g$ is also increasing in $s$, $w(s)$ is increasing. Second, note that we can interpret the variables $\frac{y(s)}{Y^*}$ and $\frac{y^*(s)}{Y^*}$ as the densities of some distribution over $s$. Thus, the desired inequality \eqref{eq: prop_FOSD} is that the expected value of $w$ is higher under the distribution associated with $y^*$ than $y$. 

The constraint that $\int_s^1 y(r) dr \leq \int_s^1 y^*(r) dr$ implies \textit{first order stochastic dominance} of the distribution whose density is $\frac{y^*}{Y^*}$ over $\frac{y}{Y^*}$. The inequality \eqref{eq: prop_FOSD} then follows because any increasing function has a higher expected value under a FOSD-higher distribution.
We have therefore concluded that in order to maximize the expected quantity procured, the integral constraint binds everywhere. Therefore, the agent's payoff must be identically equal to $V^*(s; \hat{b})$, and the envelope condition then pins down the quantities to exactly $q^*(\hat{b},s)$. Thus, the value of the relaxed program is exactly $Q^*(\hat{b})$. 

\paragraph{Step 4: For all $b > \hat{b}$, the optimal menu out of $\mathcal{C}^*$ is $C_{\hat{b}}$} 
By Step 3, it is impossible for the delegate to procure expected quantity greater than $Q^*(\hat{b})$. By single-crossing in the delegate's preferences, that means all $b > \hat{b}$ optimally procure $Q^*(\hat{b})$ units in expectation. Optimality of $C_{\hat{b}}$ for $\hat{b}$ implies that it minimizes the expected transfer subject to procuring $Q^*(\hat{b})$ units, and it is therefore optimal for all $b > \hat{b}$ as well.

\subsection*{Proof of \Propref{prop: efficient implementation general}, \Propref{prop: whom to delegate to --- info} and \Propref{prop: whom to delegate to --- ownership}}
We provide a proof for the three results, starting by constructing a standard mechanism satisfying necessary conditions for efficiency. We will use the following definitions: $q^*_i(\theta_1,\theta_2)$ is the efficient allocation for player $i$, $u_i[\theta_1,\theta_2] := \theta_i q^*_i (\theta_1,\theta_2),$ is the efficient type-dependent payoff of player $i$, and $\tau(\theta_1,\theta_2)$ is the subsidy/tax by the principal (which must be zero in \Propref{prop: efficient implementation general}). It is also useful to define the ex-ante surplus $s(\theta_1,\theta_2) = \max\{\theta_1,\theta_2\}$, and the interim surplus $S_i (\theta_i) := \int s(\theta_i,\hat\theta_{-i}) F_{-i}(d\hat\theta_{-i}) = \mu_{-i}+I_{F_{-i}}(\theta_i).$

By \theoref{theo: implementability}, efficiency can be implemented through delegation if and only if it is BIC and IIR for the delegate, and DSIC and EPIR for the agent. Let the delegate be player i. Following \citet{holmstrom1979groves}, DSIC implies the transfer paid by the agent, player $-i$, must be a Vickrey-Clarke-Groves (VCG) transfer: $t_{-i} (\hat{\theta}_1,\hat{\theta}_2) = - u_i [\hat{\theta}_1, \hat{\theta}_2] + h(\hat{\theta}_i),$ for some function $h$. 

Transfers must satisfy $t_i=-t_{-i}-\tau$, so $t_i(\hat{\theta}_1,\hat{\theta}_2) = u_i[\hat{\theta}_1,\hat{\theta}_2] - h(\hat{\theta}_i)-\tau(\hat\theta_1,\hat\theta_2)$. In expectation, delegate's payments are
$T_i(\hat{\theta}_i) := \mathbb{E}_{\tilde{\theta}_{-i}}[ t_i(\hat{\theta}_i, \tilde{\theta}_{-i}) ] = \mathbb{E}_{\tilde{\theta}_{-i}}[u_i[\hat{\theta}_i,\tilde{\theta}_{-i}] - h(\hat{\theta}_i) - \tau(\hat\theta_i,\tilde\theta_{-i}) ] $. Notice that the second term, $h(\hat\theta_i)$ is independent of $\hat{\theta}_{-i}$, so it can be pulled outside of the expectation.

BIC implies that the delegate must be compensated by the expected surplus of the agent up to a constant that does not depend on his report: $T_i(\hat{\theta}_i) = \mathbb{E}_{\tilde{\theta}_{-i}}[ -u_{-i}[\hat{\theta}_1,\tilde{\theta}_2] ] + c$.
Combining the previous two equations for $T_i$, we have:
\begin{align*}
    h(\hat{\theta}_i) &= \mathbb{E}_{\tilde{\theta}_{-i}}[u_i[\hat{\theta}_i,\tilde{\theta}_{-i}] + u_{-i}[\hat{\theta}_i,\tilde{\theta}_{-i}] ] -\mathbb{E}_{\tilde\theta_{-i}}[\tau(\hat\theta_i,\tilde\theta_{-i})]- c\\
    &= S(\hat\theta_i) - \bar\tau(\hat\theta_i)- c,
\end{align*}
where $\bar\tau(\hat\theta_i):=\mathbb{E}_{\tilde\theta_{-i}}[\tau(\hat\theta_i,\tilde\theta_{-i})]$, and the second equality follows from $$u_i[\hat{\theta}_i,\tilde{\theta}_{-i}] + u_{-i}[\hat{\theta}_i,\tilde{\theta}_{-i}]  = \theta_i q^*_i(\theta_1,\theta_2) + \theta_{-i} q^*_{-i} (\theta_1,\theta_2) = s(\theta_1,\theta_2).$$
Therefore, given subsidies, the payments are pinned down up to one constant, $c$.
\begin{align*}
    t_{-i}(\hat{\theta}_1, \hat{\theta}_2) &= - u_i[\hat{\theta}_1,\hat{\theta}_2] + S_i(\hat{\theta}_i) - \bar\tau(\hat\theta_i) - c,\\
    t_i(\hat{\theta}_1, \hat{\theta}_2) &= - t_{-i}(\hat{\theta}_1, \hat{\theta}_2)-\tau(\hat\theta_1,\hat\theta_2)\\
    &= u_i[\hat{\theta}_1,\hat{\theta}_2] - S_i(\hat{\theta}_i) + \bar\tau(\hat\theta_i) + c - \tau(\hat\theta_1,\hat\theta_2). 
\end{align*}
Given these transfers, we can calculate the ex-post payoffs of the two players:
\begin{align*}
    U_{-i}(\theta_1, \theta_2) &= u_{-i}[\theta_1,\theta_2] + u_{i}[\theta_1,\theta_2] - S_i(\theta_i) + \bar\tau(\theta_i) + c\\
    &= s(\theta_1,\theta_2) - S_i(\theta_i) + \bar\tau(\theta_i) +c,\\
    U_i(\theta_1, \theta_2) &= S(\theta_i) - \bar\tau(\theta_i) -  c + \tau(\theta_1,\theta_2).
\end{align*}
The individual rationality constraints can then be written as:
\begin{align*}
    U_{-i}(\theta_1, \theta_2) &=  s(\theta_1, \theta_2) -S_i (\theta_i) + \bar{\tau}(\theta_i)+c \geq r_{-i}  \ \theta_{-i}, \ \ \forall\theta_1,\theta_2, \tag{EPIR} \label{eq: EPIR3}\\
    \mathbb{E}_{\tilde \theta_{-i}} \left[U_i(\theta_i, \tilde{\theta}_{-i})\right] &= S_i(\theta_i) - c \geq r_i \ \theta_i,  \ \ \forall\theta_i. \tag{IIR} \label{eq: IIR3}
\end{align*}

\paragraph{Proof of \Propref{prop: efficient implementation general}}
Under this proposition, we have budget balance: $\tau=0$, and the delegate is player 1. To pin down $c$, note, from the surplus definitions, that $U_2(0,0) = c - \mu_2 \geq 0$, and $U_1(0,0) = \mu_2 - c \geq 0$, which implies $c = \mu_2$. Thus, we can specialize the individual rationality constraints to:
\begin{align*}
    U_2(\theta_1, \theta_2) &=  \max\{\theta_1,\theta_2\} - I_{F_2}(\theta_1) \geq r_2  \ \theta_2, \ \ \forall\theta_1,\theta_2, \tag{EPIR-P3} \label{eq: EPIR2}\\
    \mathbb{E}_{\tilde \theta_2} \left[U_1(\theta_1, \tilde{\theta}_2)\right] &= I_{F_2}(\theta_1) \geq r_1 \ \theta_1,  \ \ \forall\theta_1. \tag{IIR-P3} \label{eq: IIR1}
\end{align*}
We now prove the necessary and sufficient conditions. We start from \ref{eq: IIR1}, the individual rationality constraint for player 1. For sufficiency, note that $r_1=0$ implies \ref{eq: IIR1} holds, because $I_{F_2} \geq 0$. For necessity, suppose $r_1>0$. Notice that because $F_2$ is continuous and $F_2(0)=0$, there exists $x$ close enough to zero such that $F_2(x) < r_1$. Then:
$$I_{F_2} (x) - r_1 x \leq (F_2(x)-r_1) x < 0, $$
where the first inequality follows from convexity of $I_{F_2}$. This proves necessity and sufficiency of $r_1=0$ for the delegate's individual rationality constraint. 

To conclude the proof, we show that $r_2 \leq \mu_2$ is necessary and sufficient for \ref{eq: EPIR2}. Note that \ref{eq: EPIR2} is equivalent to:
$$ \min_{\theta_1} \left\{\max\{\theta_1,\theta_2\}-I_{F_2}(\theta_1) \right\} \geq r_2 \ \theta_2,  \ \ \forall\theta_2.$$
We start by showing $\min_{\theta_1} \left\{\max\{\theta_1,\theta_2\}-I_{F_2}(\theta_1) \right\} = \theta_2 - I_{F_2}(\theta_2)$. Indeed, for $\theta_1\leq\theta_2$:
\begin{align*}
    \max\{\theta_1,\theta_2\}-I_{F_2}(\theta_1) = \theta_2-I_{F_2}(\theta_1) \geq \theta_2 - I_{F_2}(\theta_2),
\end{align*}
because $I_{F_2}$ is increasing. For $\theta_1>\theta_2 $:
\begin{align*}
\begin{split}
     \max\{\theta_1,\theta_2\}-I_{F_2}(\theta_1)  = \theta_1 - I_{F_2}(\theta_1) \geq \theta_2 - I_{F_2}(\theta_2),
\end{split}
\end{align*}
since $\theta - I_{F_2} (\theta)$ is increasing. Because by choosing $\theta_1 = \theta_2$ we obtain equality, we conclude what we wanted to show. As a consequence, \ref{eq: EPIR2} is equivalent to:
\begin{align} \label{eq: EPIR2equivalent}
    \theta_2 - I_{F_2}(\theta_2) \geq r_2 \theta_2, \ \ \forall \theta_2.
\end{align}
For necessity of $r_2 \leq \mu_2$, assume otherwise---i.e. $r_2 > \mu_2$. Then, for $\theta_2=1$:
\begin{align*}
    I_{F_2}(1) = \int^1_0 F_2(v) dv = F_2(v)|^1_0 - \int^1_0 v F(dv) = 1-\mu_2 > 1-r_2,
\end{align*}
where the second equality follows from integration by parts, and the inequality from $r_2 > \mu_2$. By reorganizing the inequality above, we obtain that \ref{eq: EPIR2equivalent} fails for $\theta_2=1$. 

For sufficiency, assume $r_2 \leq \mu_2$. Then:
\begin{align*}
    (1-r_2)\theta_2 \geq (1-\mu_2) \theta_2 = I_{F_2}(1) \theta_2 = I_{F_2}(1)\theta_2 + I_{F_2}(0) (1-\theta_2) \geq I_{F_2}(\theta_2), 
\end{align*}
where the first inequality follows from $r_2 \leq \mu_2$; the second equality from $I_{F_2} (0)=0$, and the inequality from convexity of $I_{F_2}$.  \hfill $\blacksquare$

\paragraph{Proof of \Propref{prop: whom to delegate to --- info} and \Propref{prop: whom to delegate to --- ownership}} Now $\tau\neq 0$ is allowed. Start by noticing, from \ref{eq: IIR3}:
\begin{equation} \label{eq: IIR lower-bound}
c \leq \min_{x \in [0,1]}\{S_i(x) - r_i x\}    
\end{equation}
Moreover, recalling that $s(\theta_1,\theta_2) = \max\{\theta_1,\theta_2\}$, notice that:
\[r_{-i} \theta_{-i} - s(\theta_1,\theta_2)  \leq r_{-i} \theta_i - \theta_i,\]
with equality holding at $\theta_{-i}=\theta_i$. Then, \ref{eq: EPIR3} is equivalent to:
\[S_i(\theta_i) - (1-r_{-i})\theta_i - c \leq \bar{\tau}(\theta_i)\]
The minimal subsidy is achieved at the highest possible constant $c$ which, from \ref{eq: IIR lower-bound}, implies,
\[\bar{\tau}_i (\theta_i) = S_i (\theta_i) - (1-r_{-i}) \theta_i - \min_{x \in [0,1]}\{S_i(x) - r_i x\} \]
We then calculate the expected subsidy:
\[ \mathbb{E}_{\hat\theta_i}[\bar{\tau}_i(\hat{\theta_i})] = \mathbb{E}_{\hat\theta_i}[S_i(\theta_i)] - (1-r_{-i})\mu_i - \min_{x \in [0,1]}\{S_i(x) - r_i x\}. \]
Now, we prove \Propref{prop: whom to delegate to --- info}. The first term is the ex-ante surplus, which is independent of who the delegate is. Assuming $r_i=r_{-i}$ and $F_{1} \preceq_{\text{m.p.s.}} F_{2}$,
recall $S_i(x) = I_{F_{-i}}(x) + \mu_{-i}$. Because of the mean-preserving comparison, we know $I_{F_1} \leq I_{F_2}$ and $\mu_{1}=\mu_2$. Therefore, having $-i=2$ reduces the expected subsidy, and it is less expensive to delegate to $i=1$. This proves \Propref{prop: whom to delegate to --- info}.

Finally, we will prove \Propref{prop: whom to delegate to --- ownership}. Let $F:= F_i = F_{-i}$, with density $f$. Recall that $r_i + r_{-i}=1$. Then, by the argument in the previous paragraph, the choice of the delegate does not affect the first term in the expression above. Therefore, we are concerned with minimizing:
$$V(r_i) = \max_{x \in [0,1]}\{r_i (x-\mu) - I_F(x) \},  $$
where we used that $1-r_{-i}=r_i$ and $S_i (x) = I_F(x) + \mu$. Assume, without loss, that $r_1<1-r_1 = r_2$. Note that $V$ is a convex, continuous function on a compact set, so it is absolutely continuous. Given that $V(0) = 0$, the envelope theorem implies, for all $r \in [0,1]$:
$$V(r) = \int^r_0 (F^{-1}(u)-\mu) du. $$
We then obtain:

\begin{equation*}
\begin{split}
  V(1-r_1) - V(r_1) =  \int^{1-r_1}_{r_1} (F^{-1}(u)-\mu)du  = \\ (1-2r_1) \left( \frac{\int^{1-r_1}_{r_1} F^{-1}(u)du}{1-2r_1} - \int^1_0 F^{-1}(u) du \right)  = (1-2r_1)\left(\mathbb{E}[F^{-1}(X)]-\mathbb{E}[F^{-1}(Y)]\right), 
\end{split}
\end{equation*}
where the second equality follows from noticing that $\int^1_0 F^{-1}(u) du=\mu$, and $X$ and $Y$ are random variables with $X\sim U[r_1, 1-r_1]$ and $Y \sim U[0,1]$. Notice that $U[r_1,1-r_1] \preceq_{\text{m.p.s.}} U[0,1]$. 

To prove the first part of the result, first, assume $f$ is increasing, and thus $F$ is convex and $F^{-1}$ is concave. Then, because $X$ is a mean-preserving-contraction of $Y$, $V(1-r_1) - V(r_1) \geq 0$, and 1 is the optimal delegate. Symmetrically, if $f$ is decreasing and, therefore, $F^{-1}$ is convex, $V(1-r_1) - V(r_1) \leq 0$, and 2 is the optimal delegate. 

To prove the second part of the result, note once more that $V$ is convex and absolutely continuous. We can then characterize the minimal subsidy by the first-order condition $0=V'(r)= F^{-1}(r)-\mu. $ \hfill $\blacksquare$

\subsection*{Proof of \Propref{prop: resale with buyback}}
Because the optimal BIC auction is implementable in dominant strategies, \theoref{theo: implementability} guarantees there is a contracting space that achieves the same revenue. We formalize the indirect contracts in the statement of the proposition that achieve this goal. An outcome in this section is a tuple $(q_1, q_2, t_1, t_2)$, in which $q_i \in \{0,1\}$ represents whether player $i$ ends up with the good, and $t_i$ is the net transfer paid by that player. 

Define $\dw{p} = \psi^{-1}_F(0)$, $r:=\psi^{-1}_G (0)$, and $d(p)$ as in the statement. The contracting space is $\mathcal{C} = \{C^p\}_{p \in [\dw p,1]}$ such that 
$$C^p = \left\{\underbrace{(1,0,p-d(p),0),}_{\text{intermediary keeps the good}} \underbrace{(0,1,-d(p),p),}_{\text{buyer purchases}}  \ \underbrace{(0,0,p-d(p)-r,0)}_{\text{seller buys back}} \right\}.$$

We now solve for the unique (up to measure zero) equilibrium induced by this game. It is clear that it is optimal for the final consumer to buy  if and only if $p \leq \theta_2$. For the optimal choice of price by the intermediary, notice that he solves:
$$\max_p p (1-F(p)) + \theta_1 F(p) - (p-d(p)) = \max_p(\theta_1 - p) F(p) + d(p).$$
This objective is differentiable, and its derivative is:
$$ (\theta_1 - p) f(p) - F(p) + F(p) + p f(p) -\psi^{-1}_G \circ \psi_F (p) f(p)  = \left( \theta_1 - \psi^{-1}_G \circ \psi_F (p) \right) f(p), $$
which is positive for $p \leq \psi^{-1}_F \circ \psi_G (\theta_1)$, and negative otherwise. Therefore, the optimal price chosen by an intermediary of type $\theta_1$ is $p(\theta_1) :=\psi^{-1}_F \circ \psi_G (\theta_1)$. 

Because the minimal allowed price is $\dw p = \psi^{-1}_F(0)$, it must be that all types $\theta_1 \leq \psi^{-1}_G (0)$ choose $\dw p$. Moreover, for $\theta^*_1  := \psi^{-1}_G(0)$, two observations are important. (1) This intermediary is indifferent between returning and not returning the good, implying all types of the intermediary $\theta_1 < \theta^*_1=\psi^{-1}_G(0)$ return the product when they cannot sell it. (2) computing the payoff of type $\theta^*_1 = \psi^{-1}_G(0) = r$:
$$ (r - \dw p) F(\dw p) + d(\dw p) = 0,  $$
so $\psi^{-1}_G(0)$ is held to his outside option. Thus, the scheme we designed generates the same allocations as the optimal BIC auction, and gives the bidders their same interim payoffs. As a conclusion, it has the same revenue as the optimal BIC auction and is, therefore, optimal.

Finally, we show that $d(p) \in [0,p]$. That $d(p) \leq p$ is evident by its formula. To show that $d(p)\geq 0$, we start by noticing that $\psi_G \geq \psi_F$. Indeed, our gains-from-trade assumption is equivalent to:
$$\frac{G(x) - G(t)}{1-G(t)} \geq \frac{F(x) - F(t)}{1-F(t)},$$
for all $0\leq t \leq x\leq 1$. Which implies
$$\frac{1-F(x)}{1-G(x)} \geq \frac{1-F(t)}{1-G(t)}, $$
for all $0\leq t\leq x\leq 1$. In other words, the function $R(x) = \frac{1-F(x)}{1-G(x)}$ is increasing. Because it is also differentiable, we must have: 
$$0 \leq R'(x) = \frac{g(x) f(x)}{\left(1-F(x)\right)^2} \left(  \frac{1-F(x)}{f(x)} - \frac{1-G(x)}{g(x)}\right), $$
which concludes that $\psi_F(x) = x - \frac{1-F(x)}{f(x)} \leq x - \frac{1-G(x)}{g(x)} = \psi_G(x)$ for all $x \in [0,1]$.

We will now use this fact to prove $d(p) \geq 0$. Note that $d$ is differentiable with:
\begin{equation}
    \begin{split}
        d'(p) = \left( p - \psi^{-1}_G \circ \psi_F (p)\right) f(p) + F(p) \\ = \left( p - \psi_F (p)\right) f(p) + F(p) + \left(\psi_F (p) - \psi^{-1}_G \circ \psi_F (p)\right) f(p) \\ = 1 + \left(\psi_F (p) - \psi^{-1}_G \circ \psi_F (p)\right) f(p) \\ \geq 1 + \left(\psi_F (p) - \psi^{-1}_F \circ \psi_F (p)\right) f(p) \\  = 1 - (1-F(p)) \geq 0,
    \end{split}
\end{equation}
where the second equality adds and subtracts $\psi_F(p) f(p)$, the third equality uses the definition of $\psi_F$, and the inequality follows from $\psi_F \leq \psi_G$, which implies $\psi^{-1}_F \geq \psi^{-1}_G$.

Because $d$ is increasing, it is sufficient to check $d(\dw p) \geq 0$. For that:
$$d(\dw p) = (\dw p - r ) F(\dw p) = \left(\psi^{-1}_F (0) - \psi^{-1}_G(0) \right) F(\dw p) \geq 0, $$
with the inequality following again from $\psi_F \leq \psi_G$. \hfill $\blacksquare$

\bibliography{bibliography}

\section{Supplemental Appendix: Extensions} \label{sec: extensions}

In this Appendix, we extend the implementability result of \theoref{theo: implementability} in two directions. First, we allow for the delegate and the agent to have interdependent values, subject to a restriction in the information structures. Second, we consider a problem in which the delegate contracts with multiple agents simultaneously. 

\subsection*{Interdependent Values}
The setup is as in \secref{sec: model}, except that player $i$'s payoff may depend on the whole information profile $(\theta_1,\theta_2)$. We will show that our implementation result remains exactly the same to the extent that the information has a \emph{nested} structure.  

To formalize the result, we will decompose private information as $\theta_i = (\theta^c_i, \theta^p_i)$, in which $\theta^c_i$ is to be interpreted as the component of $i$'s information that affects both players' payoffs---superscript $c$ stands for common---and $\theta^p_i$ the component that affects only player $i$'s payoff. Player $i$'s payoff from outcome $x \in X$ is $u_i (x, \theta_i, \theta^c_{-i})$. We will assume that information is nested: heuristically, the agent knows more than the delegate about the common components. Formally, $\theta^c_1$ is a measurable function of $\theta_2$. That is, there exists a measurable function $\omega$ such that:
\begin{equation} \label{eq: nested information} \tag{NI}
    \theta^c_1 = \omega(\theta_2).
\end{equation}
The case in which the agent's payoff does not depend on the delegate's information is a special case of this formulation.

\begin{theorem} \label{theo: implementability with interdependence}
Suppose \eqref{eq: nested information} holds. A social choice function $Y$ is implementable through delegation if and only if it satisfies:
\begin{enumerate}[label=(\roman*)]
    \item BIC and IIR for the delegate, and
    \item DSIC and EPIR for the agent.
\end{enumerate}
\end{theorem}

\subsubsection*{Proof of \theoref{theo: implementability with interdependence}}

\subsubsection*{Only if  $\Rightarrow$} Assume that $Y$ is implementable through delegation. Let $\mathcal{C}$ be the contract space, and $\left(\sigma_1, \{\sigma_2^C,\mu^C\}_{C \in \mathcal{C}}\right)$ an associated equilibrium that implements $Y$. Because any Perfect Bayesian Equilibrium is a Bayes-Nash Equilibrium, the revelation principle implies that there exists a direct centralized mechanism in which $Y$ is implemented by truthful revelation. Formally, there exists $\mathcal{M} = \left(M, g \right)$ such that $M = \Theta_1 \times \Theta_2$, with optimal strategies $s_i (\theta_i) = \theta_i$, for $i \in \{1, 2\}$. Because this equilibrium implements $Y$, it must be the case that $g(\theta_1,\theta_2) = Y(\theta_1,\theta_2)$ for all $(\theta_1,\theta_2) \in \supp \mu_o$. Moreover, because $Y$ is deterministic, $\sigma^C_2(\theta_2)$ is deterministic for all $C \in \supp \sigma_1(\theta_1)$ and because $Y$ is implemented through delegation:
\begin{equation} \label{eq: outcome transform with 
interdependence}
 g (\theta_1, \theta_2) = \sigma^C_2(\theta_2), 
\end{equation}
for all $(\theta_1,\theta_2) \in \supp \mu_o$, $C \in \supp \sigma_1 (\theta_1)$.

Because $Y$ is implemented by $\mathcal{M}$, it must satisfy BIC and IIR for both players. In what follows we prove it also satisfies (i) DSIC and (ii) EPIR for the agent.

Fix any $\theta_1 \in \Theta_1$, and $C \in \supp \sigma_1(\theta_1)$. Because $\left(\sigma_1, \{\sigma_2^C,\mu^C\}_{C \in \mathcal{C}}\right)$ was an equilibrium in the delegated contracting game, we have, for any agent's information $\theta_2$ such that $(\theta_1,\theta_2) \in \supp \mu_o$ and any alternative outcome available to the agent, $x' \in C$:
\begin{align*}
\begin{split}
    \int_{\hat\theta_1} u_2 (\sigma^C_2(\theta_2), \theta_2, \hat{\theta}^c_1) \mu^C(d\hat{\theta}|\theta_2) = u_2 \left( \sigma^C_2 (\theta_2), \theta_2, \omega(\theta_2)  \right) \\ \geq u_2 \left(x', \theta_2, \omega(\theta_2)\right) = \int_{\hat\theta_1} u_2 (x', \theta_2, \hat{\theta}^c_1) \mu^C(d\hat{\theta}|\theta_2),
    \end{split}
\end{align*}
where the first and last inequalities follow from $u_2$ not depending on $\theta^p_1$, and from $\theta^c_1 = \omega(\theta_2)$. 

We can, in particular, choose $x' = \sigma^C_2 (\theta'_2)$ for any type of the agent with $(\theta_1,\theta'_2) \in \supp \mu_o$, and take expectation over $C$ on both sides with respect to $\sigma_1 (\theta_1)$ to obtain:
\begin{align*}
\begin{split}
u_2 \left(g(\theta_1,\theta_2),\theta_2, \omega(\theta_2) \right)     = \int_C  u_2 \left( \sigma^C_2(\theta_2), \theta_2, \omega(\theta_2)  \right)   \sigma_1(\theta_1) (dC) \\  \geq  \int_C  u_2 \left( \sigma^C_2(\theta'_2), \theta_2, \omega(\theta_2) \right)   \sigma_1(\theta_1) (dC)  = u_2 (g(\theta_1, \theta'_2), \theta_2, \omega(\theta_2)),  
\end{split}
\end{align*}
where the first and last equalities come from the equality in \ref{eq: outcome transform with interdependence}. Because $\theta_1$ was arbitrary, the inequality above shows truthful revelation is a dominant strategy for the agent in direct mechanism $\mathcal{M}$. 

Because we assume the agent can always choose $o$, the result above also implies, for any $\theta_1$:
$$u_2 (g(\theta_1,\theta_2),\theta_2, \omega(\theta_2))  \geq u_2 (o, \theta_2, \omega(\theta_2)),$$
so that truthful revelation also satisfies ex-post individual rationality. Because, by implementation, $g=Y$, we obtain that $Y$ satisfies $(i)$ and $(ii)$. 

\subsubsection*{If $\Leftarrow$} To prove the converse, let $Y$ be implementable in a centralized mechanism, and let $(i)$ and $(ii)$ hold. By the revelation principle, there is a direct mechanism $\mathcal{M} = (M,g)$ that implements $Y$ in a truthful-revealing equilibrium. Because it implements $Y$, we must have:
\begin{equation} \label{eq: outcome transform 2 with interdependence}
g(\theta_1,\theta_2) = Y(\theta_1,\theta_2).    
\end{equation}
Consider the contract space $\mathcal{C} = [C^{\theta_1}]_{\theta_1 \in \Theta_1}$, where $$C^{\theta_1} = \{y: \exists \theta_2 \text{ with } (\theta_1,\theta_2) \in \supp \mu_o \text{ and } y=Y(\theta_1,\theta_2)\}.$$ 
Given this contract space, we will show that it is a PBE for the delegation game for each type of the delegate to offer menu $C^{\theta_1}$, and, upon observing $C^{\theta_1}$, for the agent to choose contract $Y(\theta_1,\theta_2) \in C^{\theta_1}$. 

Starting with the agent's best response, because of (i), and because $\mathcal{M}$ implements $Y$, for all $\theta_1 \in \Theta_1$, $\theta_2, \theta'_2 \in \Theta_2$:
\begin{align*}
     u_2(Y(\theta_1,\theta_2),\theta_2, \omega(\theta_2)) = u_2(g(\theta_1,\theta_2),\theta_2, \omega(\theta_2)) \\ \geq  u_2(g(\theta_1,\theta'_2),\theta_2, \omega(\theta_2))   = u_2(Y(\theta_1,\theta'_2),\theta_2, \omega(\theta_2)) ,  
\end{align*}
where the equalities follow from equation \ref{eq: outcome transform 2 with interdependence}. 
Because of $(ii)$, and because $\mathcal{M}$ implements $Y$ for all $\theta_1 \in \Theta_1$, $\theta_2, \theta'_2 \in \Theta_2$:
\begin{align*}   u_2(Y(\theta_1,\theta_2),\theta_2, \omega(\theta_2)) = u_2(g(\theta_1,\theta_2), \theta_2, \omega(\theta_2)) \geq  u_2(o,\theta_2, \omega(\theta_2))  ,  
\end{align*}
so the two inequalities above show that it is optimal for the agent to reveal his type truthfully in the delegation game for any contract $C^{\theta_1}$, for any beliefs the agent might hold about the delegate.
Next, because $\mathcal{M}$ Bayes-implements $Y$ for the delegate:
\begin{align*}
    \begin{split}
   \int_{\hat\theta_2} u_1(Y(\theta_1,\hat\theta_2),\theta_1, \hat\theta^c_2)  \mu_o (\theta_1,d\hat\theta_2)   = \int_{\hat\theta_2}  u_1(g(\theta_1,\hat\theta_2),\theta_1, \hat\theta^c_2)  \mu_o (\theta_1, d\hat\theta_2) \\ \geq        \int_{\hat\theta_2}  u_1(g(\theta_1',\hat\theta_2),\theta_1, \hat\theta^c_2)  \mu_o (\theta_1, d\hat\theta_2) = \int_{\hat\theta^c_2}u_1(Y(\theta_1',\hat\theta_2),\theta_1, \hat\theta^c_2)  \mu_o (\theta_1,d\hat\theta_2),
    \end{split}
\end{align*}
Moreover, because of IIR: 
\begin{align*}
    \begin{split}
   \int_{\hat\theta_2} u_1(Y(\theta_1,\hat\theta_2),\theta_1, \hat\theta^c_2)  \mu_o (\theta_1,d\hat\theta_2)   = \int_{\hat\theta_2}  u_1(g(\theta_1,\hat\theta_2),\theta_1, \hat\theta^c_2)  \mu_o (\theta_1, d\hat\theta_2)\\ \geq  \int_{\hat\theta^c_2} u_1(o,\theta_1, \hat\theta^c_2) \mu_o (\theta_1, d\hat\theta_2),
    \end{split}
\end{align*}
implying it is an equilibrium of the delegated game for the delegate to choose the contract associated to his type, regardless of the agent's beliefs. Choose beliefs $\mu^C$ to be Bayes-consistent with the delegate's strategy on-path, and arbitrary off-path. Given that it is optimal for both players to truthfully reveal their types in the delegated game, the outcome implemented in this equilibrium is $Y (\theta_1,\theta_2)$, finishing the proof. \hfill $\blacksquare$
\subsection*{Multiple Agents}
Consider $N+1$ players, $i \in \{1,...,N+1\}$. Player 1 is the delegate, and players $2$ through $N+1$ are agents. Each player $i$ holds private information or type $\theta_i$, with $(\theta_1,...,\theta_{N+1})$ drawn jointly from $\Theta := \times_{i=1,...,N+1} \Theta_i$, according to a probability measure $\mu_o \in \Delta\Theta$. We will use the following notation: for any player $i$, let $\theta_{-i}$ be the vector of all other agents' types except for $i$. That means that the delegate's type, $\theta_1$ is not included in $\theta_{-i}$ for any $i$. Conditional on an agent's own type, we will assume that the delegate's information is independent of the other agent's types. Formally,
\begin{equation} \label{eq: agents know more about other agents} \tag{AKM}
   \theta_{-i}|\theta_i \perp \theta_1, \quad \forall i \in \{2,...,N+1\}.
\end{equation}
 Player $i$'s payoff from outcome $x$ is $u_i(x,\theta_i)$. 

We need to specify menus of contracts in a way that is consistent with multiple agents. We formalize menus as transparent mechanisms: a menu as a pair $(A^C, C)$, in which $A^C=\times_{i=2,...,N+1} A^C_i$ with $A^C_i$ being a set of actions available to player $i$; and $C: A^C \rightarrow X$ being an outcome selection as a consequence of that set of actions. A contract space is then a family $\mathcal{C}$ of menus $(A^C,C)$. We will still refer to a menu by its outcome mapping $C$.  
We will say that a social choice function $Y$ is dominant-strategy incentive compatible for player $i$ with respect to the delegate if, for all $\theta_1 \in \Theta_1$ and $\theta_i,\theta'_{i} \in \Theta_i$:
\begin{align} \label{eq: DSIC-D} \tag{DSIC-D}
    \int_{\hat\theta_{-i}} u_i (Y(\theta_1, \theta_i, \hat\theta_{-i}),\theta_i)\mu_o(d\hat\theta_{-i}|\theta_i, \theta_1) \geq \int_{\hat\theta_{-i}} u_i (Y(\theta_1, \theta'_i, \hat\theta{-i}),\theta_i) \mu_o(d\hat\theta_{-i}|\theta_i, \theta_1).
\end{align}
Similarly, we say it is ex-post individually rational for player i with respect to the delegate if, for all $\theta_1 \in \Theta_1$ and $\theta_i \in \Theta_i$:
\begin{align} \label{eq: EPIR-D} \tag{EPIR-D}
    \int_{\hat\theta_{-i}} u_i (Y(\theta_1, \theta_i, \hat\theta{-i}),\theta_i) \mu_o(d\hat\theta_{-i}|\theta_i, \theta_1) \geq u_i(o,\theta_i).
\end{align}

\begin{theorem} \label{theo: implementability with multiple agents}
Suppose \eqref{eq: agents know more about other agents} holds. A social choice function $Y$ is implementable through delegation if and only if it satisfies:
\begin{enumerate}[label=(\roman*)]
    \item BIC and IIR for the delegate, and
    \item DSIC-D and EPIR-D for all agents.
\end{enumerate}
    
\end{theorem}

\subsection*{Proof of \theoref{theo: implementability with multiple agents}}
\label{proof:theo:implementability with multiple agents}

\subsubsection*{Only if  $\Rightarrow$} Assume that $Y$ is implementable through delegation. Let $\mathcal{C}$ be the contract space, and $\left(\sigma_1, \{\sigma_i^C,\mu_i^C\}_{C \in \mathcal{C}, i=2,...,N+1}\right)$ an associated equilibrium that implements $Y$. The revelation principle implies that there exists a direct centralized mechanism in which $Y$ is implemented by truthful revelation. Formally, there exists $\mathcal{M} = \left(M, g \right)$ such that $M = \Theta$, with optimal strategies $s_i (\theta_i) = \theta_i$, for $i \in \{1, ...,N+1\}$. Because this equilibrium implements $Y$, it must be the case that $g(\theta_1,...,\theta_{N+1}) = Y(\theta_1,...,\theta_{N+1})$ for all $(\theta_1,...,\theta_{N+1}) \in \supp \mu_o$. Moreover, because $Y$ is deterministic, $\sigma^C_i(\theta_i)$ is deterministic for all $C \in \supp \sigma_1(\theta_1)$ and because $Y$ is implemented through delegation:
\begin{equation} \label{eq: outcome transform with multiple agents}
 g (\theta_1, ..., \theta_{N+1}) = C(\sigma^C_2(\theta_2),...,\sigma^C_{N+1} (\theta_{N+1})), 
\end{equation}
for all $(\theta_1,...,\theta_{N+1}) \in \supp \mu_o$, $C \in \supp \sigma_1 (\theta_1)$.

Because $Y$ is implemented by $\mathcal{M}$, it must satisfy BIC and IIR for all players. In what follows we prove it also satisfies (i) DSIC-D and (ii) EPIR-D for the agents.
Fix any $\theta_1 \in \Theta_1$, and $C \in \supp \sigma_1(\theta_1)$. Because $\left(\sigma_1, \{\sigma_i^C,\mu_i^C\}_{C \in \mathcal{C}, i=2,...,N+1}\right)$ was an equilibrium in the delegated contracting game, we have, for any agent's information $\theta_i$ such that $(\theta_1,...,\theta_{N+1}) \in \supp \mu_o$ and any alternative action available to the agent, $a' \in A_i^C$:
\begin{align*}
\begin{split}
  \int_{\hat\theta_{-i}} u_i (C(\sigma^C_i(\theta_i), \sigma^C_{-i}(\hat\theta_{-i})), \theta_i)\mu_o(d\hat\theta_{-i}|\theta_i) =\int_{\hat\theta_{-i}} u_i (C(\sigma^C_i(\theta_i), \sigma^C_{-i}(\hat\theta_{-i})), \theta_i)\mu_o(d\hat\theta_{-i}|\theta_i, \theta_1) \\  =   \int_{\hat\theta_1}\int_{\hat\theta_{-i}} u_i (C(\sigma^C_i(\theta_i), \sigma^C_{-i}(\hat\theta_{-i})), \theta_i) \mu_o(d\hat\theta_{-i}|\hat\theta_i,\hat\theta_1) \mu_i^C (d\hat\theta_1|\theta_i)  \\ \geq \int_{\hat\theta_1}\int_{\hat\theta_{-i}} u_i (C(a', \sigma^C_{-i}(\hat\theta_{-i})), \theta_i) \mu_o(d\hat\theta_{-i}|\hat\theta_i,\hat\theta_1) \mu_i^C (d\hat\theta_1|\theta_i) \\ = \int_{\hat\theta_{-i}} u_i (C(a', \sigma^C_{-i}(\hat\theta_{-i})), \theta_i)\mu_o(d\hat\theta_{-i}|\theta_i) , 
\end{split}
\end{align*}
where the first equality follows from \eqref{eq: agents know more about other agents}, the second equality follows from $\theta_1$ being payoff irrelevant in addition to \eqref{eq: agents know more about other agents}, and the last equality uses these two steps again at once. We can, in particular, choose $a' = \sigma^C_i (\theta'_i)$ for any type of agent $i$ with $(\theta_1,...,\theta'_i,...,\theta_{N+1}) \in \supp \mu_o$, and take expectation over $C$ on both sides with respect to $\sigma_1 (\theta_1)$ to obtain:
\begin{align*}
\begin{split}
 \int_{\hat\theta_{-i}} u_i (g(\theta_1,\theta_i, \hat\theta_{-i}),\theta_i) \mu_o (d\hat\theta_{-i}|\theta_i) \\ =   \int_C\int_{\hat\theta_{-i}} u_i (C(\sigma^C_i(\theta_i), \sigma^C_{-i}(\hat\theta_{-i})), \theta_i)  \mu_o(d\hat\theta_{-i}|\theta_i)\sigma_1 (dC) \\\geq \int_C\int_{\hat\theta_{-i}} u_i (C(\sigma^C_i(\theta'_i), \sigma^C_{-i}(\hat\theta_{-i})), \theta_i)  \mu_o (d\hat\theta_{-i}|\theta_i) \sigma_1 (dC) \\ = \int_{\hat\theta_{-i}} u_i (g(\theta_1,\theta'_i,\hat\theta_{-i}),\theta_i) \mu_o(d\hat\theta_{-i}|\theta_i), 
\end{split}
\end{align*}
where the first and last equalities come from \ref{eq: outcome transform with multiple agents}. Because $\theta_1$ was arbitrary, the inequality above shows truthful revelation is a dominant strategy for the agent with respect to $\theta_1$ in direct mechanism $\mathcal{M}$. 
Because we assume the agent can always choose $o$, the result above also implies, for any $\theta_1$:
$$\int_{\hat\theta_{-i}} u_i (g(\theta_1,\theta_i,\hat\theta_{-i}),\theta_i) \mu_o(d\hat\theta_{-i}|\theta_i) \geq u(o,\theta_i) ,$$
so that truthful revelation also satisfies ex-post individual rationality with respect to the delegate's type. Because, by implementation, $g=Y$, we obtain that $Y$ satisfies $(i)$ and $(ii)$.
\subsubsection*{If $\Leftarrow$} To prove the converse, let $Y$ be implementable in a centralized mechanism, and let $(i)$ and $(ii)$ hold. By the revelation principle, there is a direct mechanism $\mathcal{M} = (M,g)$ that implements $Y$ in a truthful-revealing equilibrium. Because it implements $Y$, we must have:
\begin{equation} \label{eq: outcome transform 2 with multiple agents}
g(\theta_1, ...,\theta_{N+1}) = Y(\theta_1,...,\theta_{N+1}).    
\end{equation}
Consider the contract space $\mathcal{C} = [\left(A^{\theta_1},C^{\theta_1}\right)]_{\theta_1 \in \Theta_1}$, where $A^{\theta_1}_i = \Theta_i$, for $i \in \{2,...,N+1\}$, and $$C^{\theta_1} (\theta_2,...,\theta_{N+1}) = g(\theta_1,...,\theta_{N+1}).$$ 
Given this contract space, we will show that there is a PBE for the delegation game for each type $\theta_1$ of the delegate to offer menu $C^{\theta_1}$, and, upon observing $C^{\theta_1}$, for agent $i$ to pick strategy $\sigma_i (\theta_i) = \theta_i$, for all $\theta_i \in \Theta_i$.

Starting with an agent's best response, because of (i), and because $\mathcal{M}$ implements $Y$, it must be that for all $\theta_1 \in \Theta_1$, $\theta_i, \theta'_i \in \Theta_i$:
\begin{align*}
\begin{split}
 \int_{\hat\theta_{-i}} u_i (Y(\theta_1,\theta_i,\hat\theta_{-i}),\theta_i) \mu_o(d \hat\theta_{-i}|\theta_i,\theta_1) =  \\  \int_{\hat\theta_{-i}} u_i (C^{\theta_1}(\theta_i,\hat\theta_{-i}),\theta_i) \mu_o (d\hat\theta_{-i}|\theta_i,\theta_1 ) \geq \int_{\hat\theta_{-i}} u_i (C^{\theta_1}(\theta'_i,\hat\theta_{-i}),\theta_i) \mu_o(d \hat\theta_{-i}|\theta_i,\theta_1)  \\ = \int_{\hat\theta_{-i}} u_i (Y(\theta_1,\theta'_i,\hat\theta_{-i}),\theta_i)\mu_o(d \hat\theta_{-i}|\theta_i,\theta_1)  , 
\end{split}
\end{align*}
where the equalities follow from equation \ref{eq: outcome transform 2 with multiple agents} and the definition of $C^{\theta_1}$. 

Because of $(ii)$, and because $\mathcal{M}$ implements $Y$, it must be that for all $\theta_1 \in \Theta_1$, $\theta_i, \theta'_i \in \Theta_i$:
\begin{align*}
      \int_{\hat\theta_{-i}} u_i (Y(\theta_1,\theta_i,\theta_{-i}),\theta_i) \mu_o(d \hat\theta_{-i}|\theta_i,\theta_1)  =  \\  \int_{\hat\theta_{-i}} u_i (g(\theta_1,\theta_i,\hat\theta_{-i}),\theta_i) \mu_o(d \hat\theta_{-i}|\theta_i,\theta_1) \geq u_i(o,\theta_i)  ,  
\end{align*}
so the two inequalities above show that it is optimal for the agent to reveal his type truthfully in the delegation game for any contract $C^{\theta_1}$, regardless of the beliefs the agent might hold about the delegate, as long as he believes all other agents truthfully reveal. 

Next, because $\mathcal{M}$ Bayes-implements $Y$ for the delegate:
\begin{align*}
    \begin{split}
 \int_{\hat\theta_{-1}} u_1 (Y(\theta_1,\hat\theta_{-1}),\theta_1) \mu_o(d\hat\theta_{-1}|\theta_1) = \\\int_{\hat\theta_{-1}}u_1(C^{\theta_1}(\hat\theta_{-1}),\theta_1) \mu_o(d\hat\theta_{-1}|\theta_1) \geq \int_{\hat\theta_{-1}}u_1(C^{\theta'_1}(\hat\theta_{-1}),\theta_1) \mu_o(d\hat\theta_{-1}|\theta_1) \\ =  \int_{\hat\theta_{-1}} u_1 (Y(\theta'_1,\hat\theta_{-1}),\theta_1) \mu_o(d\hat\theta_{-1}|\theta_1) ,
    \end{split}
\end{align*}
Moreover, because of IIR: 
\begin{align*}
    \begin{split}
  \int_{\hat\theta_{-1}} u_1 (Y(\theta_1,\hat\theta_{-1}),\theta_1) \mu_o(d\hat\theta_{-1}|\theta_1) = \int_{\hat\theta_{-1}}u_1(C^{\theta_1}(\hat\theta_{-1}),\theta_1) \mu_o(d\hat\theta_{-1}|\theta_1)\\ \geq u_1(o,\theta_1),
    \end{split}
\end{align*}
implying it is an equilibrium of the delegated game for the delegate to choose the contract associated to his type, as long as the agents truthfully reveal their types. Choose beliefs $\mu^C$ to follow from Bayes' rule whenever possible, and to be arbitrary off-path. Because the agents are truthful for any belief they may hold about the delegate, they  are truthful for this particular belief. Given that it is optimal for all players to truthfully reveal their types in the delegated game, the outcome implemented in this equilibrium is $Y (\theta_1,...,\theta_{N+1})$, finishing the proof. \hfill $\blacksquare$

\end{document}